\documentclass[aps,prb,twocolumn,showpacs]{revtex4-1}
\pdfoutput=1
\usepackage{graphicx}
\usepackage{dcolumn}
\usepackage{bm}
\usepackage{amsmath}
\usepackage{amssymb}
\usepackage{hyperref}
\usepackage{epstopdf}
\usepackage{caption}
\usepackage{verbatim}

\begin{document}
\title{Floquet topological systems in the vicinity of band crossings: Reservoir induced coherence and steady-state entropy production}
\author{Hossein Dehghani}
\author{Aditi Mitra}
\affiliation{Department of Physics, New York University, 4 Washington Place, New York, NY 10003, USA}
\date{\today}

\begin{abstract}
Results are presented for an open Floquet topological system represented by Dirac fermions coupled to a circularly
polarized laser and an external reservoir.
It is shown that when the separation between quasi-energy bands becomes small, and  comparable to the coupling
strength to the reservoir, the reduced density matrix in the Floquet basis, even at steady-state, has non-zero
off-diagonal elements, with the magnitude of the off-diagonal elements increasing with the strength of the coupling to the reservoir.
In contrast, the coupling to the reservoir only weakly affects the diagonal elements, hence inducing an effective coherence.
The steady-state reduced density matrix synchronizes with the periodic drive, and a Fourier analysis allows the extraction of the occupation
probabilities of the Floquet quasi-energy levels.
The lack of detailed balance at steady-state is quantified in terms of an entropy production rate, and it is shown that this
equals the heat current flowing out of the system, and into the reservoir. It is also shown that the entropy production rate
mainly depends on the off-diagonal components of the Floquet density matrix. Thus a stronger coupling to the reservoir leads to an enhanced entropy
production rate, implying a more efficient removal of heat from the system, which in turn helps the system maintain coherence.
Analytic expressions in the vicinity of the Dirac point are derived which highlights these results, and also indicates how the reservoir
may be engineered to enhance the coherence of the system.
\end{abstract}

\maketitle

\section{Introduction}
The study of periodically driven systems has seen a resurgence in recent years, appearing in many different
contexts such as: periodic drive as a means for realizing
myriad topological
phases~\cite{Oka09,Inoue10,Kitagawa10,Lindner11,Rudner13,Segev13,Hafezi13,Cayssol13,Carpentier14,Esslinger14}, as non-energy conserving examples of systems exhibiting
many-body-localization~\cite{Alessio13,Huse14,Ponte15,Lazarides15}, and as examples of systems that can support novel collective
behavior absent in static Hamiltonians~\cite{Titus14,Bukov15}. While plenty of physical insight can be gained by mapping the
time-dependent Hamiltonian into an effective time-independent Hamiltonian (the Floquet Hamiltonian)
that captures the time-evolution
over one period~\cite{Sambe73}, ultimately it is the distribution function of the particles
that needs to be properly accounted for in order to understand how much of the physics extracted simply from a spectral analysis of the
Floquet Hamiltonian, survives.

The distribution function depends on the dominant relaxation mechanisms,
i.e., whether it is
a good approximation to think of the periodically driven system to be isolated from its
surroundings so that the drive switch on protocol or the interactions between
particles determine the distribution function.~\cite{Lazarides14,Rigol14a,Dehghani14,Dehghani15a,Dehghani15b,Foster15,Cooper15b,Rosch15,Dehghani16a}
In contrast it could also be that the
the system is coupled to external leads, but is short in comparison to
electron-electron or electron-phonon (el-ph) scattering lengths, so that it is the leads that impose the occupation
probabilities~\cite{Kitagawa11,Torres14,Kundu14,Barnea16}. Finally  another commonly encountered example is
inelastic relaxation due to the system being coupled
to an external reservoir~\cite{Dehghani15a,Dehghani15b,Iadecola15a,Iadecola15b,Lindner15}.
In this paper we consider the last case discussed above, namely a periodically driven open system, where
the inelastic scattering with a reservoir determines the distribution function. We will be interested
in a circularly polarized drive which when applied to graphene, opens
up a topologically non-trivial gap at the Dirac points, inducing a Chern insulator~\cite{Oka09, Esslinger14}. Our work here
differs from our previous work on a similar system~\cite{Dehghani14,Dehghani15a,Dehghani15b} in that, we
in this work specifically consider the case where one is close to a topological phase transition,
so that the separation between quasi-energy bands are comparable to
the coupling to an external reservoir. Our past work was in the opposite limit where the quasi-energy level spacings were large as compared to coupling to
an external reservoir, and thus we were far from
any topological phase transitions.

One of the main new results in this regime is that we find an effective reservoir induced
coherence where the steady-state involves non-zero off-diagonal elements of the Floquet density matrix that grow with the strength of the coupling to the
reservoir, while the diagonal elements are relatively weakly affected. This leads to an effective reduced
density matrix $W_{\rm el}$ which becomes purer i.e., ${\rm Tr}\biggl[\left(W_{\rm el}\right)^2\biggr]$ increases with the coupling to
the reservoir.

Another new ingredient in this work is that, while it is known that generic driven dissipative
systems reach steady-states that cannot be described by an effective temperature, and hence do not
resemble a Gibbs' distribution~\cite{Hanggi2005,Kohn09,Shirai14,Dehghani14}, in this work we characterize this lack of detailed balance
by a net steady-state entropy production rate. We prove that in the steady-state where the density matrix has synchronized with the external drive,
the entropy production mainly depends on the off-diagonal components of the density matrix. Thus the more coherent the system becomes, the larger is
the entropy production rate. This is not paradoxical because the steady-state
entropy production rate equals the heat current flowing  out of the system and into the reservoir. Thus the more
efficient this flow is, the more effective the system is in maintaining coherence.

We obtain analytic expressions in the vicinity of the Dirac point, and use this to highlight the above general observations.
Our results also indicate how a reservoir can be engineered to control the entropy production rate. In
fact enhancing the latter can cause
the system to settle into more coherent and possibly even dark states where ${\rm Tr}\biggl[\left(W_{\rm el}\right)^2\biggr]=1$.

The paper is organized as follows. In section~\ref{model} we describe the model and outline the derivation of the
Floquet-Master equation, highlighting the approximations that fail once quasi-energy level spacings become small.
In section~\ref{dms} we discuss our results for the reduced density matrix and extract the occupation
probabilities of the Floquet quasi-energy levels, while in section~\ref{entropy}
we quantify the lack of detailed balance in terms of a steady-state entropy production rate.
Finally we conclude in section~\ref{concl}. Some details are relegated to the Appendices.
Appendix~\ref{appA} shows that all the components of the steady-state density matrix is synchronized with the
laser frequency. Appendix~\ref{appB} derives a general relation between the entropy production rate and the steady-state
density matrix, and shows by means of an analytic calculation at the Dirac point, that the entropy production rate is mainly controlled by the
off-diagonal components of the density matrix. Appendix~\ref{appC} provides details needed for arriving
at the analytic expressions at the Dirac point.

\section{Model}\label{model}
In the vicinity of a laser induced topological phase transition in graphene, we
may approximate graphene as Dirac fermions under a periodic drive.
This model can also alternately describe a laser applied to the 2D surface states of a 3D topological
insulator (TI)~\cite{Gedik13}.
The Hamiltonian of 2D Dirac fermions
coupled to an external circularly polarized laser, and also coupled to a reservoir of phonons is,
\begin{eqnarray}
H = H_{\rm el} + H_{\rm ph} + H_c,
\end{eqnarray}
where (setting $\hbar=1$)
\begin{eqnarray}
&&H_{\rm el}= \sum_{\vec{k}=\left[k_x,k_y\right],\sigma,\sigma'=\uparrow,\downarrow}
c_{\vec{k}\sigma}^{\dagger}\left[\vec{k} + \vec{A}(t)\right]\cdot\vec{\sigma}_{\sigma \sigma'}c_{\vec{k}\sigma'}.
\end{eqnarray}
$c^{\dagger}_{\vec{k}\sigma},c_{\vec{k}\sigma}$ are creation, annihilation operators for the Dirac fermions whose velocity $v=1$,
$\vec{\sigma}= \left[\sigma_x,\sigma_y\right]$ are the Pauli matrices which represent spins of surface states of a 3D TI,
or it represents the sub-lattice label for graphene.
$\vec{A}=\theta(t)A_0\left[\cos(\Omega t),-\sin(\Omega t)\right]$ is the circularly polarized laser which has
been suddenly switched on at time $t=0$, we will refer to this switch-on protocol as a quench. We will denote
the period of the laser as $T_{\Omega}=2\pi/\Omega$ and the temperature of the reservoir as $T$.

Here we consider coupling to 2D phonons
\begin{eqnarray}
H_{\rm ph}=\sum_{q,i=x,y}\left[\omega_{qi}b_{qi}^{\dagger}b_{qi}\right],
\end{eqnarray}
where the electron-phonon coupling is
\begin{eqnarray}
&&H_c = \sum_{\vec{k},q,\sigma,\sigma'}
c_{\vec{k}\sigma}^{\dagger}\vec{M}_{\rm ph}(k,q)\cdot\vec{\sigma}_{\sigma \sigma'}c_{\vec{k}+\vec{q}\sigma'},\\
&&\vec{M}_{\rm ph}(k,q) \!=\!\!\left[\lambda_{x,kq}\left(b_{x,q}^{\dagger}+b_{x,-q}\right), \lambda_{y,kq}\left(b_{y,q}^{\dagger}+b_{y,-q}\right)
\right].
\end{eqnarray}
There is no $\sigma_z$ term above because we have adopted a model for electron-phonon coupling consistent for graphene~\cite{Ando06}, where
the electron-phonon coupling should preserve $A$-$B$ sub-lattice symmetry. Such a symmetry is broken by terms proportional to $\sigma_z$.

In the absence of electron-phonon coupling, the problem is exactly solvable, where the time-evolution from
time $t_0$ to $t$ is
\begin{eqnarray}
|\Psi(t)\rangle = U_{\rm el}(t,t_0)|\Psi(t_0)\rangle.
\end{eqnarray}
For a spatially invariant system, the time-evolution operator factorizes
into different momenta $k$, $U_{\rm el}(t,t')=\prod_k U_{\rm el, k}(t,t')$ with
\begin{eqnarray}
U_{\rm el, k}(t,t')= \sum_{\alpha = u,d}e^{-i\epsilon_{k\alpha}(t-t')}|\phi_{k,\alpha}(t)\rangle
\langle\phi_{k\alpha}(t')|,
\end{eqnarray}
where $\epsilon_{k\alpha}$ are the quasi-energies, and $|\phi_{k\alpha}(t)\rangle$ are the time-periodic quasi-modes.
The quasi-modes and quasi-energies satisfy the following eigenvalue equation
\begin{eqnarray}
H_{\rm el}^{F}|\phi_{k,\alpha}\rangle = \epsilon_{k\alpha}|\phi_{k,\alpha}\rangle,
\end{eqnarray}
where $H_{\rm el}^{F} \equiv H_{\rm el}(t)-i\partial_{t}$, is known as the Floquet Hamiltonian.
Note that it is the
combination $|\psi_{k\alpha}\rangle \equiv e^{-i\epsilon_{k\alpha}t}|\phi_{k\alpha}(t)\rangle$ that obeys the time-dependent Schr\"odinger equation
for $H_{\rm el}(t)$, and since we are considering a two-level Hamiltonian at every momentum $k$,
there are only two distinct solutions that we label as $\alpha=u,d$.

Once the system is coupled to phonons, the problem is not exactly solvable, and in fact
on integrating out the phonon modes, it is straightforward to see that
we have an interacting electron problem.
We make progress by making certain assumptions: that the
coupling to the phonons is weak, only inter-band transitions between the quasi-energy levels are allowed (i.e, $\vec{M}_{\rm ph}(k,q)$
is peaked at $q=0$),
and that the phonons are always in thermal equilibrium at the temperature $T$.

Our assumption for neglecting intra-band transitions
is based on the fact that these occur on longer time scales than inter-band transitions. This is because the energy exchange
for the former is smaller than the latter, and also because typical electron-phonon matrix elements is stronger for optical phonons compared
with acoustic phonons~\cite{Ando02}. Thus there is an intermediate time-scale where the results obtained purely from inter-band
transitions will be valid.
In the next section, we outline the derivation of the Floquet-Master equation based on these assumptions.

\begin{figure}
\begin{center}
\includegraphics[height=8cm,width=8cm,keepaspectratio]{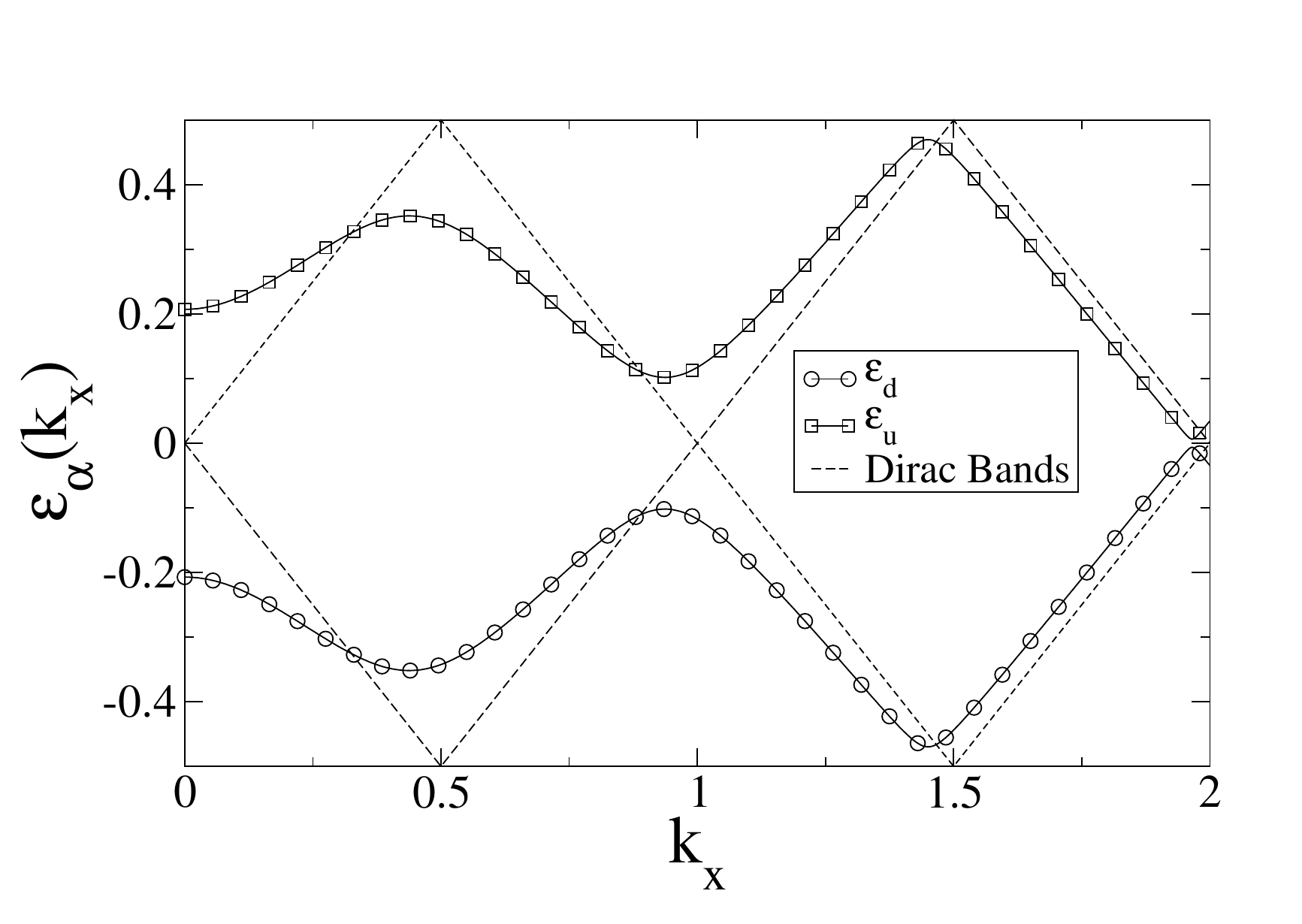}

\caption{Quasi-energy spectrum in the first FBZ for $A_0/\Omega=0.5$ along $k_y=0$, and compared with the Dirac spectrum in the absence of the laser.
We have set $\Omega=1.0$. The gap at $k=0$ is
topological and equals $\epsilon_{u}-\epsilon_d=\left(\sqrt{4 A_0^2+\Omega^2}-\Omega\right)$.}
\label{fig1}
\end{center}
\end{figure}

\subsection{Rate equation}
Let $W(t)$ be the density matrix in the Schr\"odinger picture, obeying
\begin{eqnarray}
\frac{dW(t)}{dt} = -i \left[H,W(t)\right].
\end{eqnarray}
To obtain the rate equation, it is convenient
to be in the interaction representation,
$W^{I}(t) = e^{i H_{\rm ph} t}U^{\dagger}_{\rm el}(t,0)W(t) U_{\rm el}(t,0)e^{- i H_{\rm ph} t}$.
To ${\cal O}(H_{c}^2)$, the density matrix obeys the following equation of motion
\begin{eqnarray}
&&\frac{dW^I}{dt}=-i\left[H_{c}^{I}(t),W^I(t_0)\right]\nonumber\\
&&-\int_{t_0}^tdt'\left[H_{c}^{I}(t),\left[H_{c}^{I}(t'),W^{I}(t')\right]\right],
\end{eqnarray}
where
$H_{c}^{I}$ is in the interaction representation.
We assume that at the initial time $t_0$, the electrons and phonons are uncoupled so that
$W(t_0) = W_{\rm el,0}(t_0)\otimes W_{\rm ph}(t_0)$. We assume that
initially, before the laser has been switched on, the electrons are in the ground state of Dirac fermions, while
the phonons are in thermal equilibrium at temperature $T$.

Thus for a laser quench at $t=0$,
\begin{eqnarray}
W_{\rm el}^{0}(t)= |\Psi(t)\rangle\langle \Psi(t)|=\prod_kW_{\rm el,k}^{0},
\end{eqnarray} where
\begin{eqnarray}
\!\!W_{\rm el,k}^{0}(t)=\!\!\!\! \sum_{\alpha,\beta=\pm}\!\!\!e^{-i(\epsilon_{k \alpha}-\epsilon_{k \beta})t}|\phi_{k\alpha}(t)\rangle
\langle\phi_{k\beta}(t)|
\rho_{k,\alpha\beta}^{I,\rm quench},
\end{eqnarray}
with
\begin{eqnarray}
&&\rho_{k,\alpha\beta}^{I,\rm quench}= \langle \phi_{k\alpha}(0)|\psi_{{\rm in},k} \rangle\langle\psi_{{\rm in},k}
| \phi_{k\beta}(0)\rangle,\\
&&|\psi_{{\rm in},k}\rangle = \frac{1}{\sqrt{2}}\begin{pmatrix}-(k_x-ik_y)/k\\1\end{pmatrix}.
\end{eqnarray}
The above is simply stating that for a quench, the occupation of the Floquet-levels ($\rho_{k,\alpha\beta}^{I,\rm quench}$)
are simply given by their overlap with the ground-state $|\psi_{{\rm in},k}\rangle$ of the Dirac fermions.

Defining the electron reduced
density matrix as the one obtained from tracing over the phonons,
$W_{\rm el} = {\rm Tr}_{\rm ph}W$, and noting that $H_c$ being linear in the phonon operators,
the trace vanishes, we need to solve,
\begin{eqnarray}
\frac{dW_{\rm el}^I}{dt}=-{\rm Tr}_{\rm ph}\int_{t_0}^tdt'\left[H_{c}^{I}(t),\left[H_{c}^{I}(t'),W^{I}(t')\right]\right].
\end{eqnarray}
We assume that the phonons are an ideal reservoir and stay in equilibrium with temperature $T$. In that case $W^I(t) = W_{\rm el}^I(t)\otimes
e^{-H_{\rm ph}/T}/{\rm Tr}\left[ e^{-H_{\rm ph}/T}\right]$ (we set $k_B=1$).

The most general form of the reduced density matrix
for the electrons is
\begin{eqnarray}
W_{\rm el}^I(t)  =\prod_k \sum_{\alpha\beta}\rho_{k,\alpha \beta}^{I}(t) |\phi_{k,\alpha}(t)\rangle\langle\phi_{k,\beta}(t)|,
\end{eqnarray}
where in the absence of phonons, $\rho_{k,\alpha \beta}^I=\rho_{k,\alpha\beta}^{I,\rm quench }$ and are time-independent
in the interaction representation. With phonons, $\rho_{k,\alpha\beta}^{I}(t)$ are time-dependent. We make the Markov assumption that the reservoir correlation
times is very fast as compared to the time-scale over which $\rho_{k}^{I}$ vary~\cite{Hanggi2005,Kohn09}. This allows us to pull the $\rho(t')$ out
of the integral above,
leading to the Floquet-Master equation. Since eventually one is interested in the density matrix in the Schr\"odinger representation, here we present the Floquet-Master equation in the Schr\"odinger picture. Before doing this let us define our notation for the density matrix in the Schr\"odinger picture
\begin{eqnarray}
W_{\rm el}(t)  =\prod_k \sum_{\alpha\beta}\rho_{k,\alpha \beta}^{S}(t) |\phi_{k,\alpha}(t)\rangle\langle\phi_{k,\beta}(t)|,
\end{eqnarray}
where $\rho_{k,\alpha \beta}^{S}=\rho_{k,\alpha\beta}^{I} e^{-i(\epsilon_{k\alpha}-\epsilon_{k\beta})t}$. Since in all of our future results we use
$\rho_{k,\alpha \beta}^{S}$, we will drop the superscript $S$ after this. Using this notation the Floquet-Master equation becomes
\begin{eqnarray}
&&\dot{\rho}_{k,\alpha\beta}(t) + i \left(\epsilon_{k\alpha}-\epsilon_{k\beta}\right) \rho_{k,\alpha\beta}(t)=\nonumber\\
&&-\sum_{\delta\gamma}\biggl[R^k_{\alpha \delta,\delta \gamma}(t)\rho_{k,\gamma \beta}(t)
+ \rho_{k,\alpha \gamma}(t)R^{k*}_{\beta\delta,\delta \gamma}(t)\nonumber\\
&&- \rho_{k,\delta \gamma}(t)R^{k*}_{\delta \alpha,\beta\gamma}(t)-\rho_{k,\gamma \delta}(t)
R^k_{\delta\beta,\alpha\gamma}(t)\biggr],
\label{rateo1gen}
\end{eqnarray}
where $R$ is the transition or rate matrix. Assuming a uniform phonon density of states $\nu$ and denoting $N$ as the Bose distribution function
at temperature $T$, we can use the fact that the rate matrix $R$ has the periodicity of the laser to Fourier expand it,
\begin{eqnarray}
&&R_{\alpha\beta,\alpha'\beta'}^k(t)=\sum_{n_1,n_2}e^{i(n_2-n_1)\Omega t}R^{n_2,n_1}_{\alpha\beta,\alpha'\beta'},\\
&&R^{n_2,n_1}_{\alpha\beta,\alpha'\beta'}=\biggl[\left(1+N\left[\epsilon_{k\beta'}-\epsilon_{k\alpha'}+n_1\Omega\right]\right)\nonumber\\
&&\times \theta(\epsilon_{k\beta'}-\epsilon_{k\alpha'}+n_1\Omega)\nonumber\\
&&+N\left[-\epsilon_{k\beta'}+\epsilon_{k\alpha'}-n_1\Omega\right]\theta(-\epsilon_{k\beta'}+\epsilon_{k\alpha'}-n_1\Omega)\biggr]\nonumber\\
&&\times \biggl[\left(C_{1\alpha\beta}^{n_2}C_{1\alpha'\beta'}^{-n_1}+ C_{2\alpha\beta}^{n_2}C_{2\alpha'\beta'}^{-n_1}\right)\nu\left(\lambda_x^2-
\lambda_y^2\right)\nonumber\\
&&+\left(C_{1\alpha\beta}^{n_2}C_{2\alpha'\beta'}^{-n_1}+ C_{2\alpha\beta}^{n_2}C_{1\alpha'\beta'}^{-n_1}\right)\nu\left(\lambda_x^2+\lambda_y^2\right)\biggr].
\label{Rdef}
\end{eqnarray}
where $C_{1,2}^n$ are the Fourier transform of the following matrix elements,
\begin{eqnarray}
\langle \phi_{k\alpha}(t)|c_{k\uparrow}^{\dagger}c_{k\downarrow}|\phi_{k\beta}(t)\rangle = \sum_{n}e^{in \Omega t}C_{1k\alpha\beta}^n,\label{for1}\\
\langle \phi_{k\alpha}(t)|c_{k\downarrow}^{\dagger}c_{k\uparrow}|\phi_{k\beta}(t)\rangle = \sum_{n}e^{in \Omega t}C_{2k\alpha\beta}^n.\label{for2}
\end{eqnarray}
Due to our choice of the electron-phonon coupling, any phonon absorption and emission takes place via spin/pseudo-spin flips as can be
seen explicitly from the structure of the matrix-elements $C_{1,2}$.
It is also useful to note that under complex-conjugation we have,
\begin{eqnarray}
&&R_{\alpha\beta,\alpha'\beta'}^{k*}(t)= \sum_{n_1,n_2}e^{i(n_2-n_1)\Omega t}\biggl[R^{-n_2,-n_1}_{\alpha\beta,\alpha'\beta'}\biggr]^*.
\end{eqnarray}

From Eq.~\eqref{Rdef} it is clear that $R^{n_2,n_1}_{\alpha\beta,\alpha'\beta'}$ is the Fermi-Golden rule rate for making a transition
from quasi-energy level $\epsilon_{k\alpha'}$ to $\epsilon_{k\beta'}+n_1\Omega$ by phonon absorption or emission. Since $\alpha,\beta=u,d$,
we see that this rate includes processes that change the electronic state ($\alpha\neq \beta$) as well as Floquet-Umpklapp processes
where the electron state remains the same $\alpha=\beta$, but the system absorbs or emits phonons at energy $n_1\Omega$. As
we shall later discuss, the latter
processes are particularly important for achieving reservoir induced coherence as they take away the energy being supplied by the
periodic drive without changing the electronic state (since $\alpha=\beta$).

If we had relaxed the assumption
that the phonons were in equilibrium at temperature $T$, then the functions
$N(\epsilon)=\langle b^{\dagger}_{\epsilon}b_{\epsilon}\rangle,1+N(\epsilon)=\langle b_{\epsilon} b^{\dagger}_{\epsilon}\rangle$ entering the rates
would have been unknown, and a separate kinetic equation would have to be written for them, leading to a complex electron-phonon coupled
Boltzmann equation.

We now mention some additional commonly made assumptions to further simplify the rate equation~\eqref{rateo1gen}. When the separation between
quasi-energy levels is large as compared to the coupling to the reservoir, the off-diagonal matrix elements of
the density matrix in the Floquet basis become small, and can be neglected~\cite{Kohn09}. In this case, the
steady-state solution for the density matrix, even though it bears little resemblance to a Gibbs' distribution
(unless of course the frequency of the laser is large as compared to the electron band-width), is
independent of the electron-reservoir coupling.~\cite{Dehghani14,Dehghani15a}
The simplest way to see this lack of dependence on the coupling is to notice that
when the off-diagonal elements of the density matrix are zero in Eq.~\eqref{rateo1gen},
the diagonal-elements at steady-state are given by the ratio of some combination of the rates $R$.
Since each $R$ depend on the electron-phonon coupling by being proportional to $\nu\lambda^2_{x,y}$,
the coupling dependence falls off in the steady-state result for the density matrix when $\lambda_x = \lambda_y$.
When $\lambda_x\neq \lambda_y$, the result depends on the coupling only via the asymmetry ratio $\lambda_x^2/\lambda_y^2$.
The resulting distribution at steady-state is isotropic in momentum space when $\lambda_x=\lambda_y$, while it
can be anisotropic in momentum space for $\lambda_x\neq \lambda_y$.

In addition, if the coupling to the reservoir is small
in comparison to the drive frequency $\Omega$, then the steady-state density matrix varies slowly over one cycle of the drive.
In this case, one may make the ``modified rotating wave approximation''~\cite{Kohn09, Hanggi2005} which involves replacing the
scattering rates $R(t)$ by their average over one cycle. We call this case the time-averaged Floquet-Master equation.
In what follows, we will not make either of the above assumptions, i.e., we will
retain the off-diagonal component of the density matrix and keep the
full time-dependence of $R(t)$, i.e, we will solve Eq.~\eqref{rateo1gen} directly.

In the remaining paper, we present all results for isotropic electron-phonon couplings ($\lambda_x=\lambda_y=\lambda$). Since the
resulting distribution is isotropic in momentum space, we will present results along $k_y=0$.
We also set the phonon density of states to be uniform and equal to $\nu=1$. Thus we assume that we have non-zero phonon density of states at all
relevant energy-scales. We supplement these results with analytic results at $k=0$
that highlight how the energy-dependence of the phonon density of states affect the steady-state.

We present our numerical results for phonon temperature $T=0.1\Omega$ and three different reservoir coupling strengths $\lambda/\Omega =0.08,0.16,0.2$
and in those cases where the results are weakly dependent on $\lambda$, we only present the results for $\lambda/\Omega =0.08,0.2$.
We also fix the laser amplitude to $A_0/\Omega=0.5$. The corresponding quasi-energy spectra
plotted within the 1st Floquet Brillouin zone (FBZ) ($|\omega|<\Omega/2$) is shown in Fig.~\ref{fig1}, showing
the topological gap opening at $k=0$. This curve can be compared with the dashed-line in this plot which illustrates an ideal
repeated Dirac band structure in the Floquet zone. For weak laser frequencies ($A_0/\Omega \ll 1$), we expect resonances at $|k|\sim n\Omega/2$,
with this condition shifting as the laser amplitude increases. The resonances are reflected by the narrowing of the quasi-energy
level spacings at these $k$ points, and in what follows, special attention will be given to the occupation probabilities
in the vicinity of these points, showing that the off-diagonal components of the density matrix also become larger here.

The master equation, Eq.~\eqref{rateo1gen}, is a linear ordinary differential equation (ODE) of order one with time dependent coefficients.
From ODE theory we know that such differential equations have closed solutions. While our results are based on numerical simulations of
these equations, and an analytic solution at $k=0$,
there are some general features in the steady-state of these solutions which we discuss in Appendix~\ref{appA}.
In particular one can show that in the steady-state which is reached after an initial transient whose duration is
controlled by $\lambda^2\nu$, the density matrix synchronizes with the laser field. Thus the only oscillations remaining in the system
are those with frequency $\Omega$ and its multiples.
This can at first seem counter-intuitive especially for off-diagonal components of the density matrix because one may naively
expect from the term $i(\epsilon_{k\alpha}-\epsilon_{k\beta})\rho_{k,\alpha\beta}$ on the left hand side of the master equation
Eq.~\eqref{rateo1gen}, that the off-diagonal components must oscillate with the frequency $(\epsilon_{k\alpha}-\epsilon_{k\beta})$.
As explained in the Appendix~\ref{appA}, while this is valid initially, in the steady-state these oscillations decay due to the presence of the
electron-phonon coupling terms on the right hand side of the master equation.
Therefore in the steady-state we can expand the density matrix via a Fourier series
\begin{eqnarray}
\rho_{k,\alpha\beta}^{\rm SS}(t) = \sum_m e^{im\Omega t}\rho^m_{k, \alpha\beta},
\label{Fourier1}
\end{eqnarray}
where the superscript $\rm{SS}$, denotes the steady-state. Above one may interpret $|\rho_{k,\alpha\alpha}^m|$
as the occupation of the $m$-th Floquet state of quasi-energy $\epsilon_{k\alpha}-m\Omega$, and $|\rho_{k,\alpha\beta}^m|$
is the probability of being in a coherent superposition of
the quasi-energy levels $\epsilon_{k\beta}$ and $\epsilon_{k\alpha}-m\Omega$. Furthermore, as we show later,
the entropy production rate depends on $\partial_{t}{\rho}_{k,\alpha\beta}^{\rm SS}$. Thus the Fourier expansion coefficients
$|\rho_{k,\alpha\beta}^m|$ find physical significance in terms of the entropy produced
and heat released to the environment. In what follows we will give explicit results for the parameters mentioned above.

\section{Results for the reduced density matrix}\label{dms}

\begin{figure}
\begin{center}
\includegraphics[height=8cm,width=8cm,keepaspectratio]{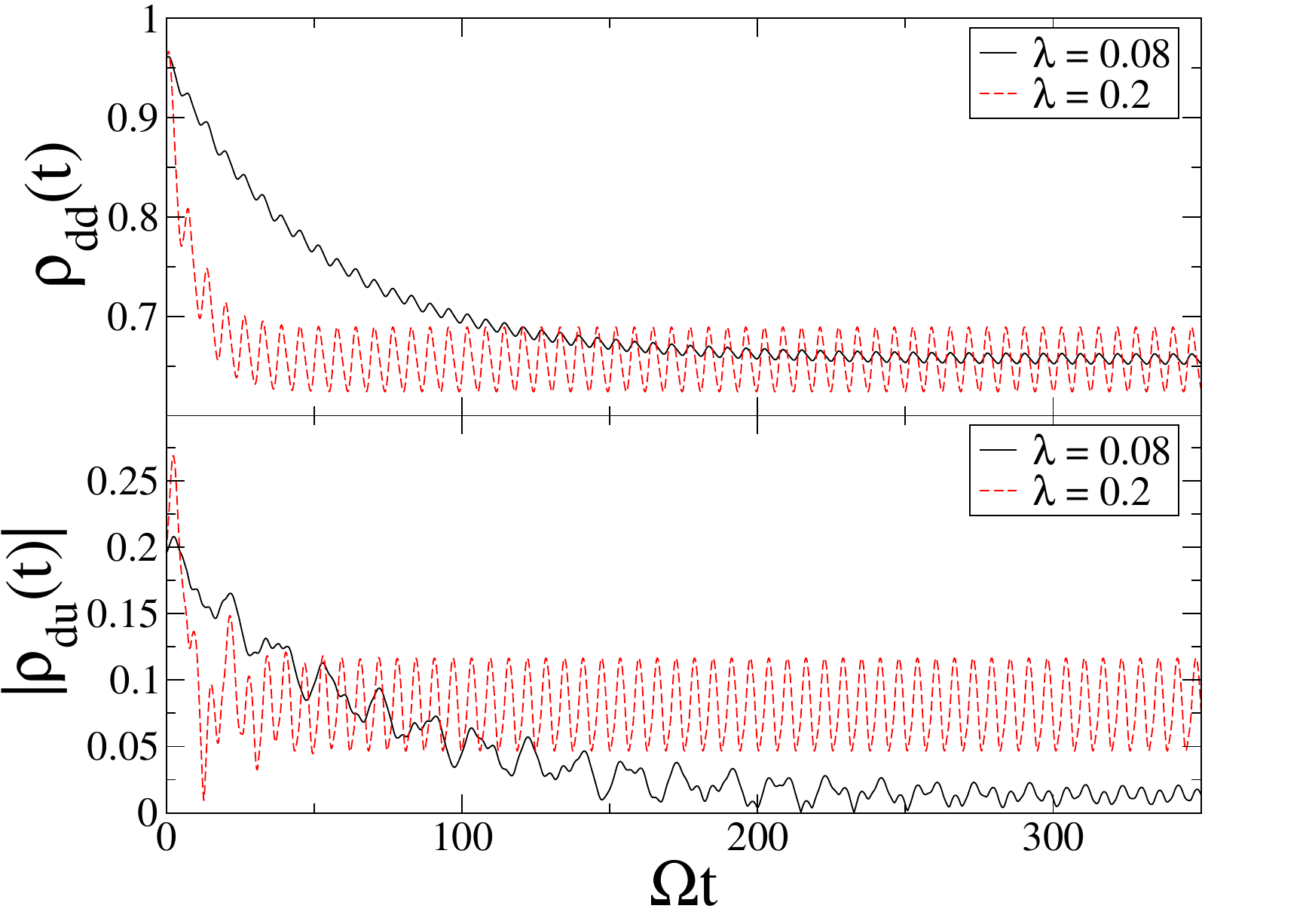}
\caption{(Color online): Time-evolution of the diagonal (upper-panel) and off-diagonal (lower-panel) components of the reduced density matrix at momentum
$k_x=0.3,k_y=0$, for electron-phonon
coupling strengths of $\lambda/\Omega=0.08,0.2$, and phonons at temperature $T=0.1\Omega$. We have set $\Omega=1.0$. Increasing $\lambda/\Omega$
decreases the time to reach steady-state,
increases the amplitude of the steady-state oscillations, and increases the magnitude of the time-averaged $\rho_{du}$, while only weakly
affecting the magnitude of the time-averaged $\rho_{dd}$.}
\label{fig2}
\end{center}
\end{figure}

\begin{figure}
\begin{center}
\includegraphics[height=8cm,width=8cm,keepaspectratio]{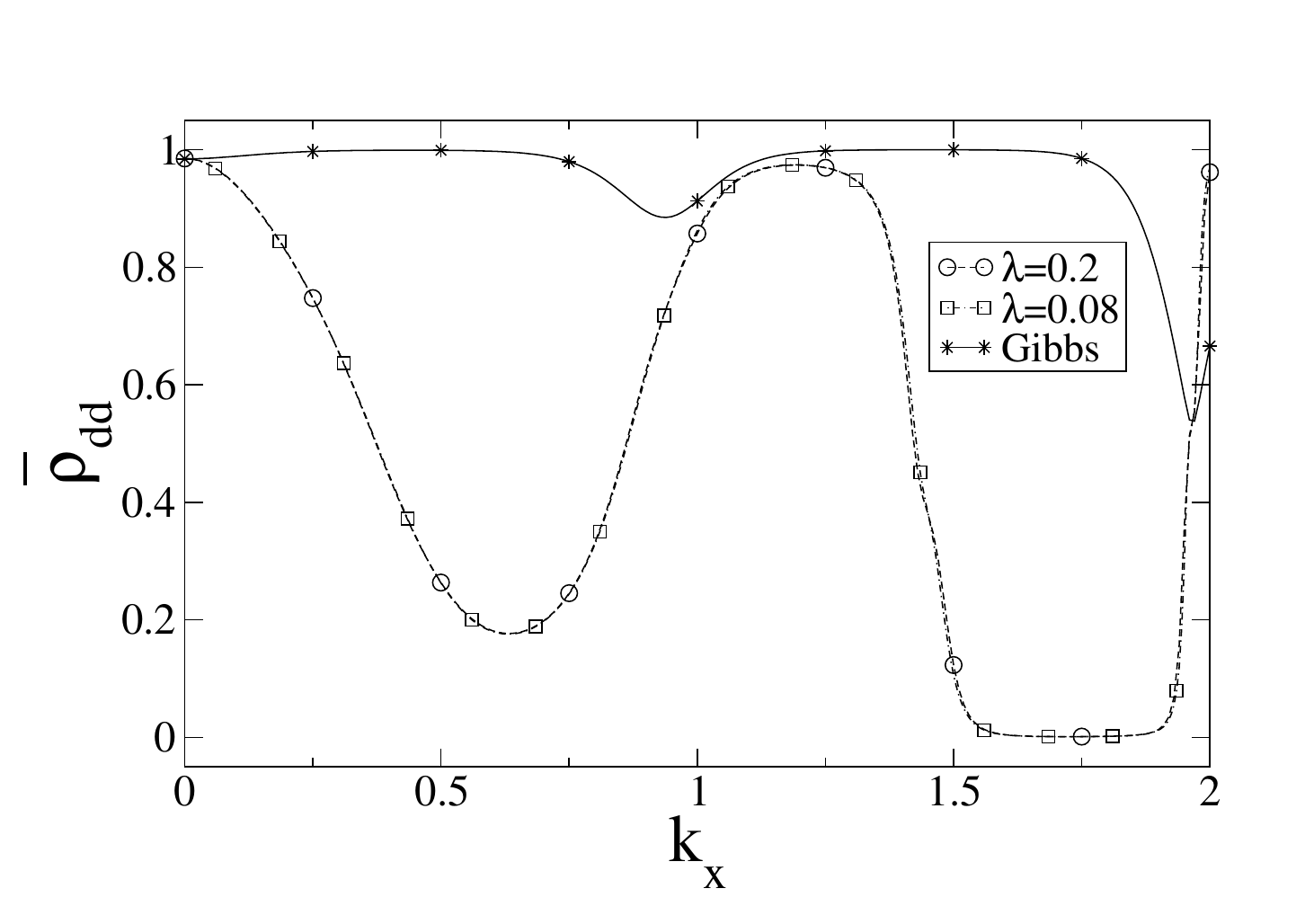}
\caption{Time-averaged diagonal component of the density matrix for the "down" Floquet level ($\bar{\rho}_{dd}$) for $\lambda/\Omega = 0.08,0.2$,
and reservoir temperature  $T=0.1\Omega$.  This is compared with the Gibbs distribution $\rho_{dd}^{\rm Gibbs}= e^{{(\epsilon_u-\epsilon_d)}/T}
/(1+e^{{(\epsilon_u-\epsilon_d)}/T})$.
We have set $\Omega=1.0,k_y=0$. $\lambda$ only weakly affects $\bar{\rho}_{dd}$, however the latter is far from a Gibbs state.}
\label{fig3}
\end{center}
\end{figure}

Fig.~\ref{fig2} shows how the diagonal and off-diagonal components of the reduced density matrix evolve in time for two different electron-phonon coupling
strengths and for a particular  momentum, we have chosen $k_x=0.3,k_y=0$. Quite generically,
the steady-state is periodic with frequency $\Omega$, and the larger the electron-phonon coupling $\lambda$,
the larger is the magnitude of oscillations in the steady-state. These oscillations originate from the the non-zero Fourier harmonics of the $R$ matrix. 
In fact as we show in Appendix~\ref{appA}, in a Floquet-Master equation with a time-averaged $R$, the steady-state oscillations of 
$\rho_{k,\alpha\beta}$ disappear.

The time-averaged value of the reduced density matrix is such that the diagonal
component is not very sensitive to $\lambda$ as can be seen in Fig.~\ref{fig3}. In this figure, in addition to the populations in the presence of
the electron-phonon coupling, there is another curve depicting the thermal distribution, denoted by $\rho_{k,dd}^{\rm{Gibbs}}=1-\rho_{k,uu}^{\rm{Gibbs}}$
and given by
$\rho_{k,uu}^{\rm{Gibbs}} = \big(1+{e^{{(\epsilon_{ku}-\epsilon_{kd})/T}}}\big)^{-1}$.
The figure shows that $\rho_{k,dd}^{\rm{Gibbs}}$ is always close to $1$, except around the resonances $k\sim\Omega, 2\Omega$ when the quasi-energy level separation
becomes small.
Moreover the Gibbs distribution
and the reservoir-induced distributions can be very different. In Ref.~\onlinecite{Dehghani14} we showed how  the
reservoir induced distribution and the Gibbs distribution approach each other at small momenta ($k\ll \Omega$)
as $A_{0}/\Omega$ becomes small, {\sl i.e.},
in the highly off-resonant case.
\begin{figure}
\begin{center}
\begin{tabular}{lll}
\includegraphics[height=8cm,width=8cm,keepaspectratio]{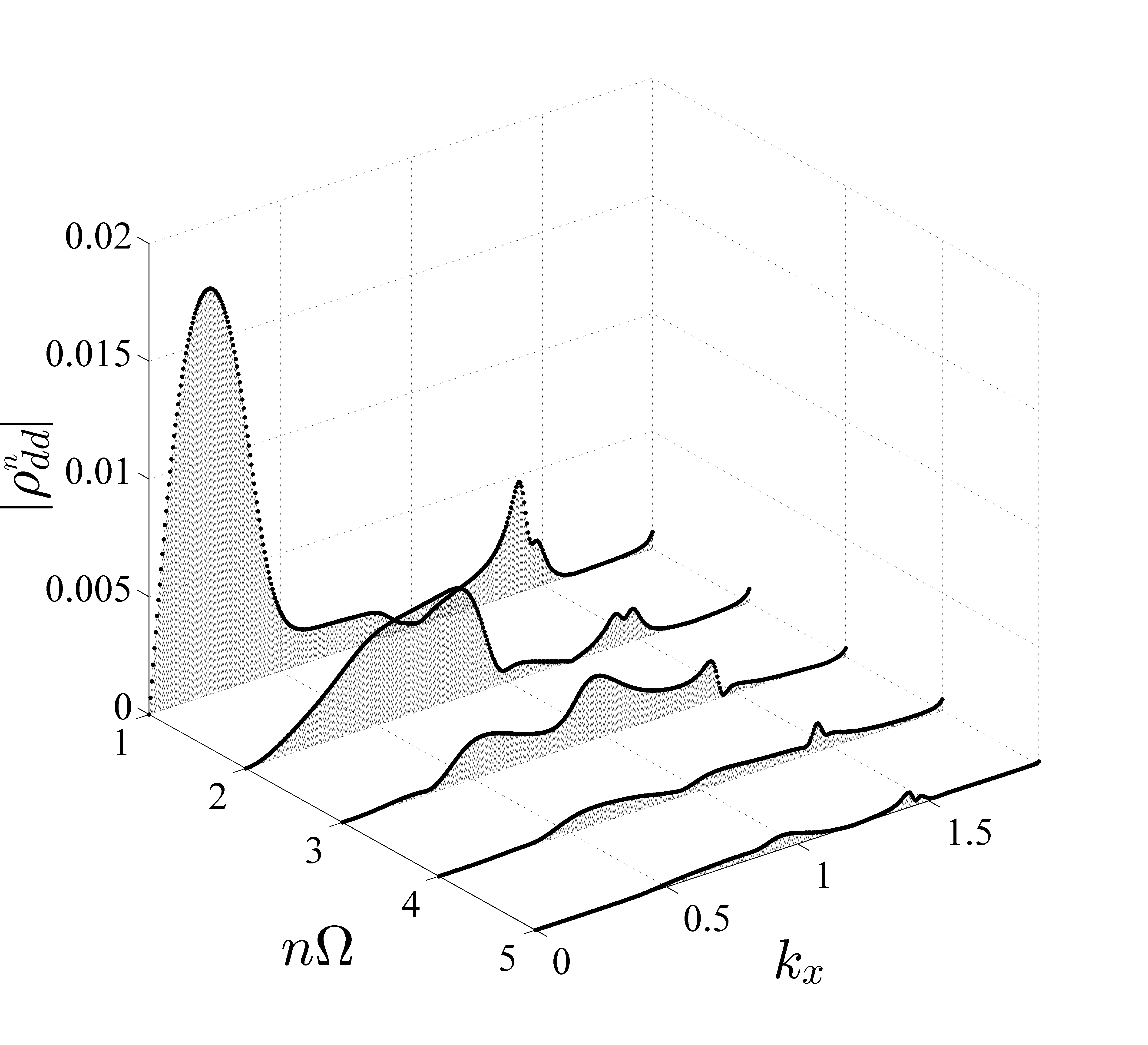}\\
(a)\\
\includegraphics[height=8cm,width=8cm,keepaspectratio]{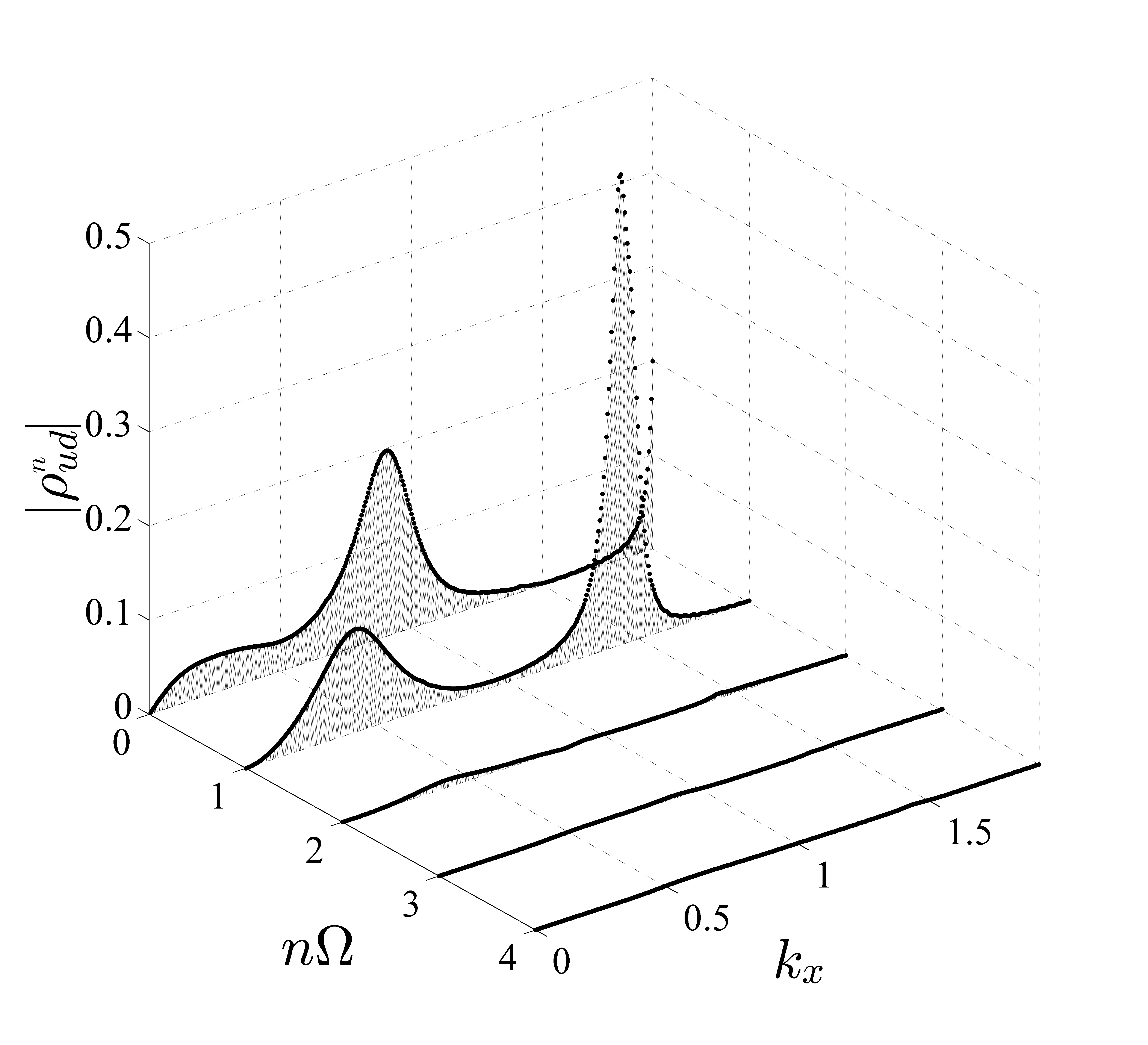}\\
(b)
\end{tabular}
\caption{Fourier transform of the (a) diagonal $\rho_{dd}^n$ and (b) off-diagonal $\rho_{ud}^n$ components of the
reduced density matrix at steady-state, for $\lambda/\Omega=0.2,k_y=0$ and reservoir temperature $T=0.1\Omega$.
The $|\rho_{dd}^n|$ are the occupation probability of the $\epsilon_d-n\Omega$ quasi-energy level,
while $|\rho_{ud}^n|$ the probability of being in a coherent superposition of quasi-energy levels $\epsilon_d$
and $\epsilon_u-n\Omega$. We have set $\Omega=1.0$. Note that $\rho_{dd}^{n=0}$ is not shown as it is already plotted
in Fig.~\ref{fig3}.
}
\label{fig4}
\end{center}
\end{figure}

\begin{figure}
\begin{center}
\begin{tabular}{lll}

\includegraphics[height=8cm,width=8cm,keepaspectratio]{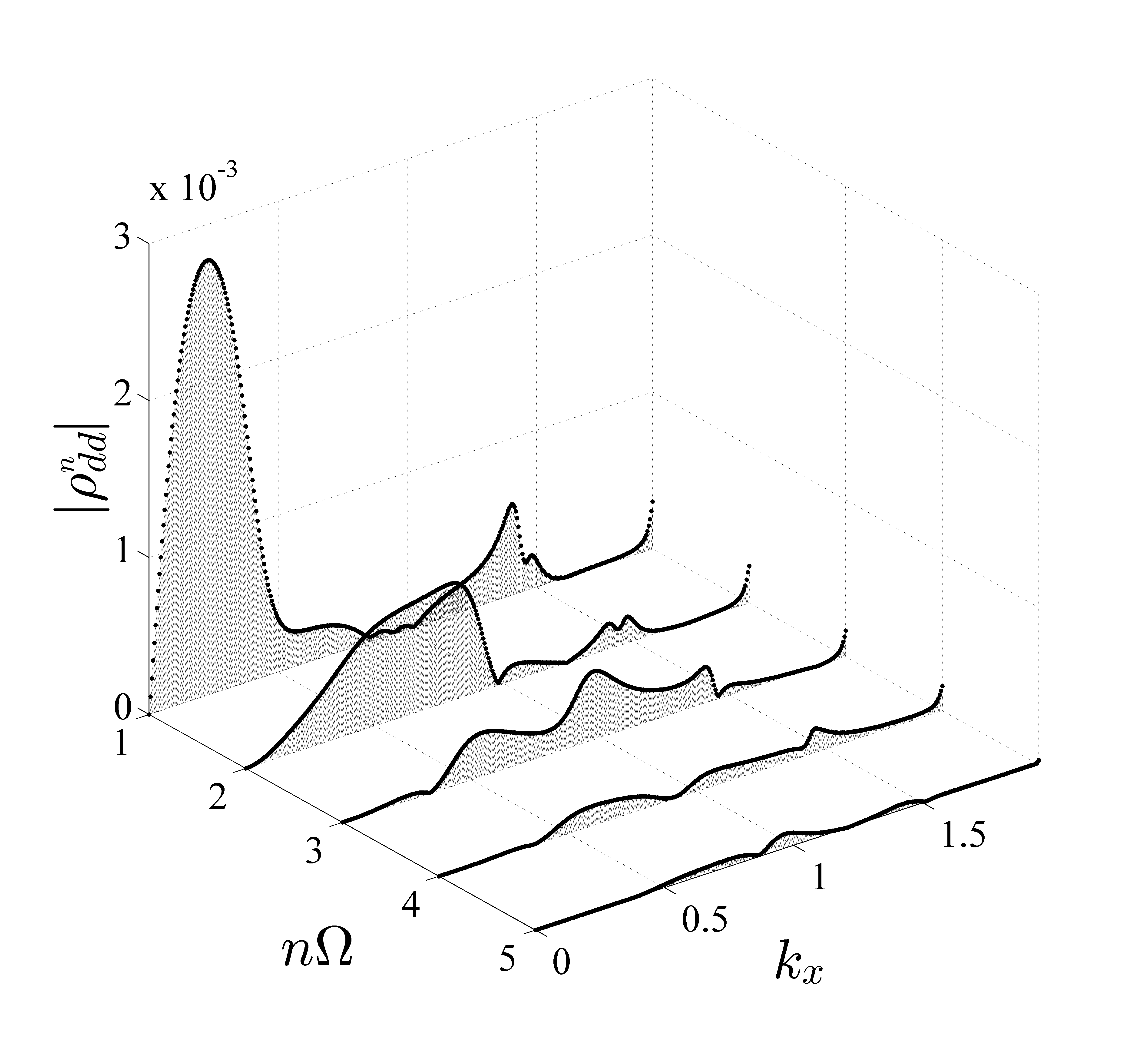}\\
(a)\\
\includegraphics[height=8cm,width=8cm,keepaspectratio]{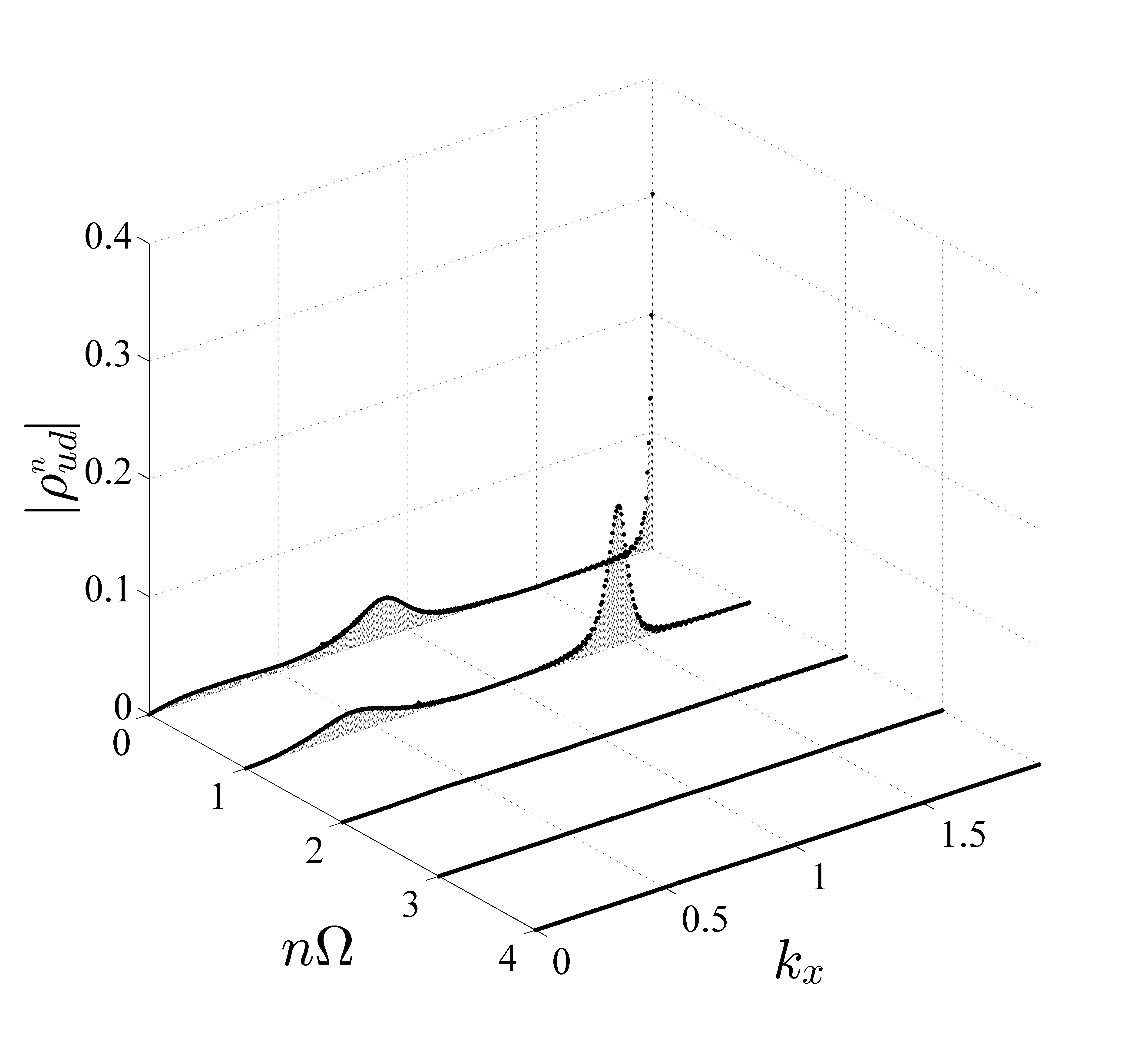}\\
(b)
\end{tabular}
\caption{
Fourier transform of the (a) diagonal $\rho_{dd}^n$ and (b) off-diagonal $\rho_{ud}^n$ components of the
reduced density matrix at steady-state, for $\lambda/\Omega=0.08,k_y=0$ and reservoir temperature $T=0.1\Omega$. We have set $\Omega=1.0$.
Note that $\rho_{dd}^{n=0}$ is not shown as it is already plotted
in Fig.~\ref{fig3}.}
\label{fig5}
\end{center}
\end{figure}

\begin{figure}
\begin{center}
\includegraphics[height=8cm,width=8cm,keepaspectratio]{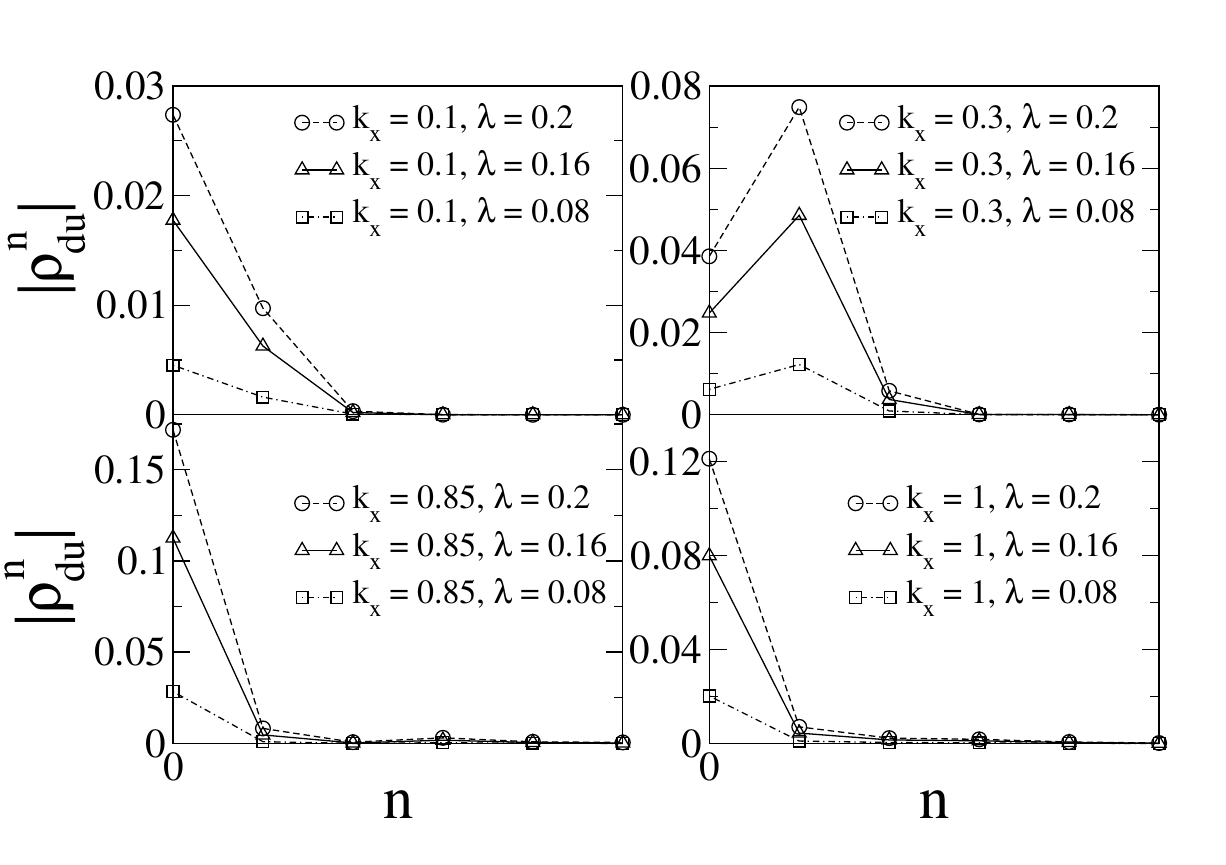}
\caption{Comparison of the off-diagonal Floquet occupation probabilities for three different coupling strengths to the reservoir:
$\lambda/\Omega=0.08,0.16,0.2$ and $T=0.1\Omega,k_y=0$. We have set $\Omega=1.0$.}
\label{fig6}
\end{center}
\end{figure}

Since the steady-state density matrix is periodic in $\Omega$, we can Fourier transform it according to Eq.~\eqref{Fourier1}, and the results
for $\lambda=0.2$ are shown in Fig.~\ref{fig4}, and those for the smaller coupling of $\lambda=0.08$ are shown on Fig.~\ref{fig5}. We
only show positive harmonics, as $|\rho_{\alpha\beta}^{n}|$ are symmetric under $n$ to $-n$. One finds that as one
approaches the resonance condition $k_x\sim \Omega/2,\Omega, 3\Omega/2 \ldots $, higher and higher harmonics of the density matrix are excited,
with their magnitude also increasing with the coupling $\lambda$ to the reservoir.
The reservoir dependence in $|\rho_{ud}^{n}|$ is shown more clearly in Fig.~\ref{fig6} where a direct comparison has been made
between three different couplings to the reservoir, and for some special values of $k$ that are close to resonance.
While in these plots for most momenta, the amplitude of higher Fourier modes are weaker than the lower Fourier modes,
around resonances it is possible that this trend becomes reversed. For instance Fig.~\ref{fig6} shows that for $k_x = 0.3$, which is close to a
resonance, $|\rho^{1}_{du}|$ is greater than $|\rho^{0}_{du}|$. This enhancement of $n=1$ photon processes is also visible in $\rho_{du}^n$ 
plotted in Fig.~\ref{fig4} and Fig.~\ref{fig5} for $k_x\sim 0.5,1.5$.

\begin{figure}
\begin{center}
\includegraphics[height=8cm,width=8cm,keepaspectratio]{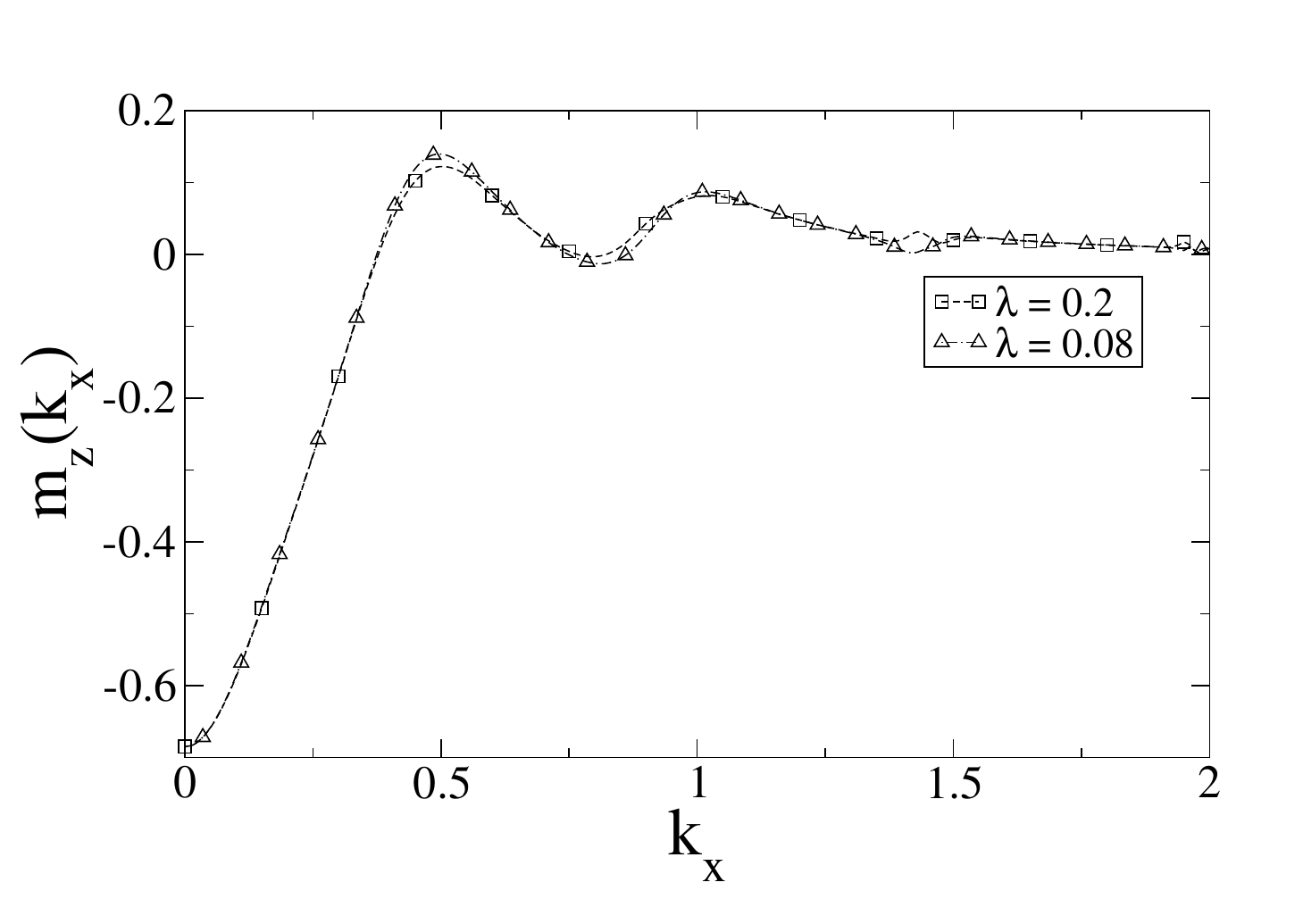}
\caption{Time-averaged spin density at steady-state along $k_y=0$ and for reservoir temperature $T=0.1\Omega$. We have set $\Omega=1.0$.
The differences for different $\lambda$ albeit small, are most pronounced near resonance ($k_x\sim 0.5, 1$).}
\label{fig7}
\end{center}
\end{figure}

In order to understand how measurable quantities are affected, in Fig.~\ref{fig7} we plot the time-averaged spin-density defined as
$m_z(k)=\frac{1}{T_{\Omega}}\int_0^{T_\Omega}dt m_{z}(k,t)$ where,
\begin{eqnarray}
&&m_z(k,t) = {\rm Tr}\biggl[W_{\rm el}(t)\sum_{\sigma}\sigma c_{k\sigma}^{\dagger}c_{k\sigma}\biggr]\nonumber\\
&&=\sum_{\alpha,\beta=u,d}\biggl[\rho_{k,\alpha\beta}(t) \sum_{\sigma}\sigma \langle \phi_{k\beta}(t)|c_{k\sigma}^{\dagger}c_{k\sigma}|\phi_{k\alpha}(t)\rangle\biggr].
\end{eqnarray}
Note that this quantity is zero in the absence of the drive, and a non-zero value of $m_z$ is a consequence of the broken time-reversal
symmetry under the influence of the circularly polarized laser.
The momentum resolved spin density can be measured in spin-resolved ARPES~\cite{Gedik13}
(Angle Resolved Photo-Emission Spectroscopy), an experimental technique
capable of observing the distribution of electrons in a spin resolved manner.
This method for example has been used to show
spin-momentum locking of the surface states of 3D topological insulators.~\cite{Hsieh09}
In an earlier paper we discussed~\cite{Dehghani14} the spin texture for the Floquet-Dirac system
when only diagonal components of the density matrix are kept. With the inclusion of off-diagonal terms as done here,
we find that a qualitatively new feature
is a dependence of the results on the coupling strength to a reservoir. While this dependence is weak for the parameters we have chosen, near a topological
phase transition, where the levels come even closer together, the reservoir dependence of $m_z$ will become more enhanced.

As the coupling strength to the reservoir is increased, the diagonal
component of the density matrix is not significantly affected, but the off-diagonal component is strongly affected,
and increases with $\lambda$. We therefore coin the term ``reservoir induced coherence''. This is unusual as typically in most open systems,
the reduced density matrix is strongly dephased by coupling to a reservoir, making it approach a diagonal ensemble.
Of course, one may always choose a time-dependent basis where the density matrix looks effectively diagonal.
What we find here is not a trivial basis dependent effect because the quantity which measures the purity of the density matrix ${\rm Tr}
\biggl[\left(W_{\rm el}\right)^2\biggr]$
increases in our model as coupling to the reservoir is increased. For a pure system ${\rm Tr}\biggl[\left(W_{\rm el}\right)^2\biggr]=1$,
while for a mixed state  ${\rm Tr}\biggl[\left(W_{\rm el}\right)^2\biggr]<1$. For our system ${\rm Tr}\biggl[\left(W_{\rm el}\right)^2\biggr]=2|\rho_{k,du}|^2 + 1 +
2 \rho_{k,dd}\left(\rho_{k,dd}-1\right)$. Thus it is clear that if $\rho_{k,dd}$ is only weakly affected by $\lambda$ (as shown in Fig.~\ref{fig3})
while $|\rho_{k,du}|$ strongly increases with $\lambda$, it will lead to a purer state.
In fact as shown in Appendix~\ref{appC}, an analytic calculation at the Dirac point ($k=0$) can be done. Here we find that for a weak laser field ($A_0/\Omega \ll 1$), and for a reservoir temperature that is small as compared to the quasi-energy level spacing ($T \ll \sqrt{4A_0^2 +\Omega^2}-\Omega\sim 2 A_0^2/\Omega$),
\begin{eqnarray}
&&\rho_{dd} = 1 + {\cal O} \biggl(\frac{A_0}{\Omega}{\rm Re}\left[\rho_{du}\right]\biggr)\nonumber\\
&&=1 + {\cal O} \biggl(\frac{\lambda^4\nu^2 A_{0}^2}{\Omega^4}\biggr).
\end{eqnarray}
Thus the above confirms our observation of a very weak dependence of the diagonal component of the density matrix on the electron-reservoir
coupling strength. In contrast as shown in Appendix~\ref{appC}, the off-diagonal component (in particular, its imaginary part)
is  ${\cal O}\left(\lambda^2\nu A_0/\Omega^2\right)$,
and depends much more sensitively on the coupling to the reservoir.

It is interesting to note that such reservoir induced coherence has also been predicted in other driven-dissipative
systems~\cite{Taylor05,Su13,breuerbook},
where they are encountered when two or more two-level systems are coupled to the same reservoir. In such a situation, the reservoir can
induce an effective entanglement between the two level systems. In our model similar physics is at play. Even though at each $k$, we have
a single two-level system corresponding to the sub-lattice of graphene or to the spin on the surface of a TI, the periodic drive introduces
effectively many levels, the Floquet levels. Now in our Floquet-Master equation, all the Floquet levels at a given $k$ are coupled to the
same phonon reservoir, resulting in a similar reservoir
induced entanglement or coherence between the Floquet levels. In the next section we explicitly show that this enhanced coherence arises because
the reservoir can absorb the excess entropy in the system, especially when there are reservoir density of states
at multiples of the driving frequency.

\begin{figure}
\begin{center}
\includegraphics[height=8cm,width=8cm,keepaspectratio]{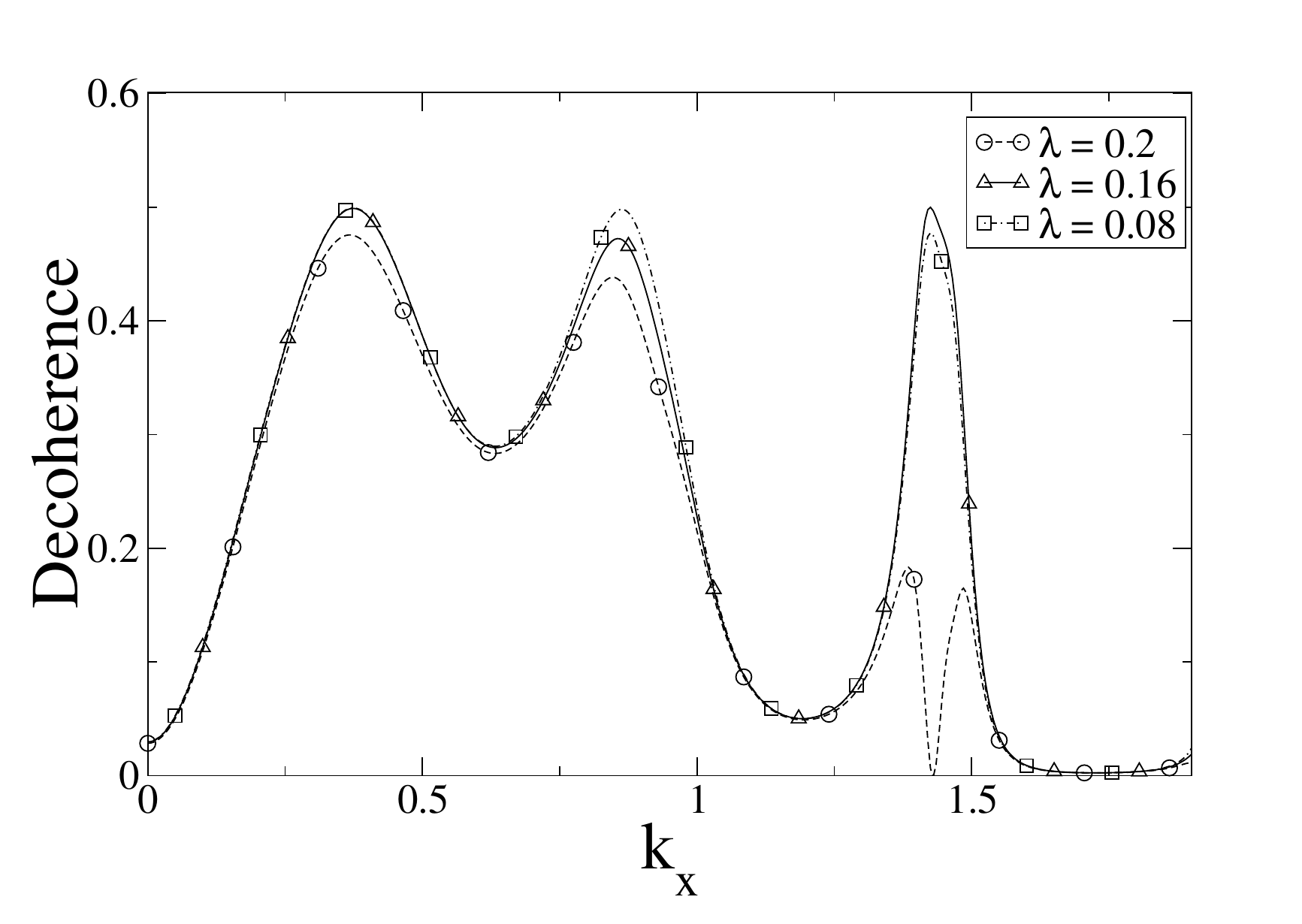}
\caption{Plot of steady-state decoherence measured as the time-average of $1-{\rm Tr}\biggl[\left(W_{\rm el}\right)^2\biggr]$, for $\lambda/\Omega=0.08,0.16,0.2$, with $\Omega=1.0,T=0.1\Omega$. The decoherence decreases as the coupling to the reservoir increases, with indication of a dark state around
$k_x\sim 1.42$.
}
\label{fig8}
\end{center}
\end{figure}

It is convenient to define a decoherence measure for the steady-state by $1-{\rm Tr}\bigl[\left(W_{\rm el}^{\rm SS}\right)^2\bigr]$.
This decoherence measure vanishes for the pure state, and increases as the density matrix becomes more mixed. The decoherence
is plotted in Fig.~\ref{fig8} for $\lambda/\Omega=0.08,0.16, 0.2$ and shows that the reservoir induced coherence increases with $\lambda$,
and is enhanced closer to resonances, with the possibility of complete coherence around $k_x\sim 1.42$. Such coherent states appearing out of dissipative
coupling to the reservoir are also known as dark states~\cite{breuerbook}.

\section{Steady-state entropy production rate and coherence}\label{entropy}

The steady-state generally does not coincide with a Gibbs' distribution, where for the latter one would expect the following
for the time-averaged distribution
function,
$\rho_{k,uu}^{\rm Gibbs}=\big(e^{(\epsilon_{ku}-\epsilon_{kd})/T}+1\big)^{-1}, \rho_{k,dd}^{\rm Gibbs}=1-\rho_{k,uu}^{\rm Gibbs}$, with
$T$ being the temperature of the reservoir.
One may quantify this lack of detailed balance in the system in terms of an entropy production rate~\cite{Zia07,Lutz11}.
Below we derive a general expression for it, and then apply it to our system.

For any time-dependent Hamiltonian, and for a time-evolution from some initial time $t=0$ to $t=\tau$,
the first law of thermodynamics states that the mean work $\langle w \rangle$ performed during this interval,
the mean heat $\langle Q\rangle$ exchanged with a reservoir at temperature $T=\beta^{-1}$, and
the change in the internal energy $\Delta U$ of the system are related as,
\begin{eqnarray}
\Delta U = \langle w \rangle + \langle Q\rangle,\label{dU}
\end{eqnarray}
where $\Delta U = U_{\tau}-U_0$. The above is simply stating energy conservation.

The entropy at any given
time is $s_t = -{\rm Tr}\biggl[W_{\rm el}(t)\ln W_{\rm el}(t)\biggr]$.
The second law of thermodynamics states that the entropy is always greater than or equal to the
heat exchanged with the reservoir. The amount by which the entropy is larger than the heat exchanged ($\Delta s-\beta \langle Q\rangle$)
can be thought of as a net entropy production which is always non-negative. Thus
the mean entropy production over the time interval
from $t=0,\tau$ may be defined as
\begin{eqnarray}
 \Sigma  = \Delta s -\beta \langle Q \rangle.
\end{eqnarray}
Using Eq.~\eqref{dU}, this implies,
\begin{eqnarray}
 \Sigma  = \Delta s + \beta \langle w \rangle- \beta  \Delta U.\label{sig1}
\end{eqnarray}
Now we give a quick derivation of the entropy production rate $\partial_t\Sigma $,
as outlined in Ref.~\onlinecite{Lutz11}.

We start with microscopic definitions of the internal energy and the mean work in terms of the density matrix. The internal energy is given by
\begin{eqnarray}
U_t = {\rm Tr}\biggl[W_{\rm el}(t) H_{\rm el}(t)\biggr].\end{eqnarray}
The average work done $\langle w \rangle $ during a time interval from $0$ to $\tau$ may be written as~\cite{Lutz11}
\begin{eqnarray}
\langle w\rangle = \int_0^{\tau}dt {\rm Tr}\biggl[W_{\rm el}(t) \partial_t H_{\rm el}(t)\biggr].\label{meanWork}
\end{eqnarray}
Let us also define a density matrix representing an ideal Gibbs' state at time $t$,
\begin{eqnarray}
W_{\rm el}^{{\rm eq}}(t)=\frac{e^{-\beta H_{\rm el}(t)}}{Z_t}; Z_t={\rm Tr }\biggl[e^{-\beta H_{\rm el}(t)}\biggr].\label{equilDensity}
\end{eqnarray}
The entropy production rate can be obtained from differentiating Eq.~\eqref{sig1} with time.
Note that the rate of change of the entropy is given by,
\begin{eqnarray}
\dot{s}(t) = -{\rm Tr}\biggl[\dot{W}_{\rm el}(t)\ln W_{\rm el}(t)\biggr],\label{ds2}
\end{eqnarray}
where we have used that $\partial_t{\rm Tr}[W_{\rm el}(t)] ={\rm Tr}[\partial_tW_{\rm el}(t)]= 0$.
In the above and in the remainder, an over-dot denotes a time
derivative as in $\dot{W}_{\rm el}(t) = \partial_{t}{W}_{\rm el}(t)$.

The change in the internal energy $U$ in time, has two contributions, one from the change in the distribution
function $W_{\rm el}$ with time, and the second from changes to the Hamiltonian $H_{\rm el}(t)$ with time. Identifying the latter as the rate at
which work is done,
\begin{eqnarray}
\beta \partial_t\left[\langle w\rangle-U\right]= -\beta {\rm Tr}\biggl[\dot{W}_{\rm el}(t)H_{\rm el}(t)\biggr].\label{dw2}
\end{eqnarray}
One immediately obtains from Eq.~\eqref{sig1},~\eqref{ds2},~\eqref{dw2},
\begin{eqnarray}
&&\dot{\Sigma}= -{\rm Tr}\biggl[\dot{W}_{\rm el}\ln \frac{W_{\rm el}(t)}{W_{\rm el}^{{\rm eq}}(t)}\biggr].\label{St}
\end{eqnarray}
Eq.~\eqref{St}
shows that the entropy production rate is zero when either $\dot{W}_{\rm el}(t)=0$ (i.e., the system has reached a time-independent steady-state)
and/or detailed balance is obeyed in that the distribution function equals the Gibbs' distribution, $W_{\rm el}(t)= W_{\rm el}^{{\rm eq}}(t)$.
We will now show that the steady-state in a Floquet system where the density matrix has synchronized with the laser, is characterized
by a net entropy production rate, which also increases with coupling to the reservoir.

A useful quantity is the entropy production rate averaged over one cycle of the laser. Since
at steady-state, due to synchronization, $W_{\rm el}^{\rm SS}(t) = W_{\rm el}^{\rm SS}(t+T_{\Omega})$, the time average vanishes
\begin{eqnarray}
\frac{1}{T_{\Omega}}\int_0^{T_{\Omega}} dt {\rm Tr}\biggl[\dot{W}_{\rm el}^{\rm SS}\ln W_{\rm el}^{\rm SS}(t) \biggr] =0.
\end{eqnarray}
Thus the time-averaged entropy production rate simplifies to
\begin{eqnarray}
\overline{\dot{\Sigma}^{\rm SS}} = -\frac{\beta}{T_{\Omega}}\int_0^{T_{\Omega}}dt{\rm Tr}\biggl[\dot{W}_{\rm el}^{\rm SS}H_{\rm el}(t)\biggr],\label{eprate1}
\label{srate}
\end{eqnarray}
where an overline denotes time averaging over one cycle.

To clarify the meaning of the entropy production rate in the steady-state, we note that the time averaged rate of
change of the entropy vanishes due to the time periodicity of the density matrix,
\begin{eqnarray}
\overline{\dot{s}^{\rm SS}}&&= -\frac{1}{T_{\Omega}}\int_0^{T_{\Omega}}dt{\rm Tr}\biggl[\dot{W}_{\rm el}^{\rm SS}{\rm ln}W_{\rm el}^{\rm SS}(t)\biggr]\notag
\\&&=0.\label{eprate2}
\label{SSentropy}
\end{eqnarray}
Similarly one can show that $\overline{\dot{U}^{\rm SS}} = 0$.
Therefore from Eq.~\eqref{sig1}, in the steady-state one finds
\begin{eqnarray}
\overline{\dot{\Sigma}^{\rm SS}} = \beta\overline{\langle \dot{w}^{\rm SS}\rangle},
\end{eqnarray}
or equivalently
\begin{eqnarray}
\overline{\dot{\Sigma}^{\rm SS}} = - \beta\overline{\langle \dot{Q}^{\rm SS}\rangle}.\label{patialSSHeat}
\end{eqnarray}
Therefore in the steady-state, a non-vanishing entropy production rate is equivalent to the heat current flowing out of the system and into the
reservoir.  Since according to the second law of thermodynamics the entropy production is always non-negative, when this quantity is nonzero,
it implies that the work performed on the system cannot be absorbed in the internal energy of the system.
This energy is converted into heat and flows out of the system towards the reservoir.

We have proved in Appendix~\ref{appB} that in the steady-state where the density matrix synchronizes with the periodic drive,
the entropy production rate can be expressed as follows in terms of the components of the density matrix and the
quasi-modes,
\begin{eqnarray}
&&\overline{\dot{\Sigma}^{\rm SS}}= -\beta\sum_{\alpha,\beta}\overline{\langle\phi_{k,\beta}(t)|\dot{\phi}_{k,\alpha}(t)\rangle\bigl(\epsilon_{k\beta}-\epsilon_{k\alpha}+i\partial_{t}\bigr)\rho_{k, \alpha\beta}^{\rm SS}}.\notag\\
\label{entRateSSFinalm}
\end{eqnarray}
We note that the above quantity is real, and also independent of the gauge, i.e., it does not depend on the arbitrariness in choosing the quasi-modes
and quasi-energies (see App.~\ref{appB} for a discussion of this point).
In addition we find that the result depends weakly on the diagonal components of the density matrix and it is mainly controlled by
the off-diagonal components of the density-matrix. While we can show this analytically at $k=0$, for non-zero $k$, this is an
observation from our numerical simulations.
Thus, after ignoring the oscillations of the diagonal elements of the density matrix, and for a two-level system, one finds,
\begin{eqnarray}
&&\overline{\dot{\Sigma}^{\rm SS}} \simeq 2\beta\nonumber\\
&&\times \!\!\mbox{Re}\!\Big[\overline{\left(\left(\epsilon_{ku} - \epsilon_{kd}\right)\rho_{k,du}^{\rm SS} + i\dot{\rho}_{k,du}^{\rm SS}\right)\langle\dot{\phi}{}_{k,u}(t)|\phi_{k,d}(t)\rangle}\Big].
\label{TwoLevelSigmaDot}
\end{eqnarray}

It is interesting to note that if we had made the commonly employed Floquet-Markov approximation~\cite{Kohn09} which involves replacing the
rates by their time-averaged values, then by construction $\dot{\rho}_{k,du}^{\rm SS}=0$, and in that case the above expression for the
entropy-production rate would have been gauge-dependent, and hence unphysical. Thus by keeping the full time-dependence of the rates, we
have a correct measure of how non-Gibbsian the resulting steady-state is.

Now let us turn to an analytic study for the entropy production rate at $k=0$. In App.~\ref{appC} we showed that for a certain convenient
gauge choice $\partial_t{\rho_{k=0,du}^{SS}}=0$. Moreover independent of the gauge $\partial_t{\rho}_{k=0, dd}^{SS}=0$.
In this case, one finds that the entropy-production rate is
\begin{eqnarray}
\overline{\dot{\Sigma}^{\rm SS}}(k=0)= 2\beta A_0\Omega {\rm Im}\bigl[\rho_{k=0,du}^{\rm SS}\bigr].
\end{eqnarray}
Thus at the Dirac point, the entropy production rate is proportional to the magnitude of the off-diagonal component of the
density matrix, and an electric field due to the laser given by the combination $A_0\Omega$.
An analytic expression may be obtained for the off-diagonal component. While originally to simplify the numerical calculations, we assumed a
model with energy independent electron-phonon coupling and density of states, at $k = 0$ we can restore the energy dependence
of these parameters.
When $A_0/\Omega\ll 1$, we find
\begin{eqnarray}
{\rm Im}\bigl[\rho_{k=0,du}\bigr] = \frac{\left(2\lambda_{\Omega}^2\nu_{\Omega}\right)A_0}{\Omega^2 + \left(2\lambda_{-}^2\nu_{-}\right)^2
\left[1+2N_{-}\right]^2},\label{offd3a}
\end{eqnarray}
where the subscripts ${\Omega}$ and $-$ in $\lambda$ and $\nu$ denote the value of these quantities at energy
$\Omega$ and energy equal to the topological gap $\Omega_{-} = \Delta - \Omega \approx 2A_{0}^2/\Omega$, respectively.
Therefore by tracing back the origins of the interactions, one can detect the corresponding microscopic processes which produce
coherence. For $A_0/\Omega\ll 1$, the main processes responsible for creation of off-diagonal components and therefore coherence are the
Floquet-Umklapp processes during which electrons are allowed to absorb or emit phonons with energy $\Omega$ from or into the reservoir.
This explains the presence of $\lambda_{\Omega}^2\nu_{\Omega}$ in the numerator of Eq.~\eqref{offd3a}.

By detecting the dominant interaction one can engineer the reservoir so as to increase coherence.
Here, this is realized by enhancing the density of states of phonons or the coupling constant at energy $\Omega$. Thus as
the drive pumps energy at $\Omega$, the system can stay coherent by releasing this energy into the reservoir.

The denominator of Eq.~\eqref{offd3a} also shows that the dissipation due to the reservoir phonons at the gap energy
$\Omega_{-}$ measured by the coupling $\lambda_{-}^2\nu_{-}$, is enhanced by the Bose factor $(1+2N_{-})$.
Note that the laser induced gap at the Dirac point is of the order of $\rm{100meV}$ in most currently accessible
setups~\cite{Gedik13}. At temperatures that are small as compared to this gap ($\beta A_0^2/\Omega \gg 1$),
\begin{eqnarray}
{\rm Im}\bigl[\rho_{k=0,du}\bigr]\sim
\nu_{\Omega} \lambda_{\Omega}^2 \frac{A_0}{\Omega^2},
\end{eqnarray}
so that,
\begin{eqnarray}
&&\overline{\dot{\Sigma}^{\rm SS}}(k=0,A_0/\Omega \ll 1)\sim \beta \left(A_0\Omega\right)\left(\nu_{\Omega}
\frac{\lambda^2_{\Omega}}{\Omega}\right)\frac{A_0}{\Omega},
\end{eqnarray}
or equivalently from Eq.~\eqref{SSentropy}
\begin{eqnarray}
\overline{\dot{Q}^{\rm SS}}(k=0,A_0/\Omega \ll 1)\sim - \left(A_0\Omega\right)\left(\nu_{\Omega} \frac{\lambda_{\Omega}^2}{\Omega}\right)\frac{A_0}{\Omega}.
\end{eqnarray}
Thus for low field amplitudes and low temperatures,
the heat production rate increases with the effective electric field $A_0\Omega$, and is constant as a function of the temperature.

\begin{figure}
\begin{center}
\includegraphics[height=8cm,width=8cm,keepaspectratio]{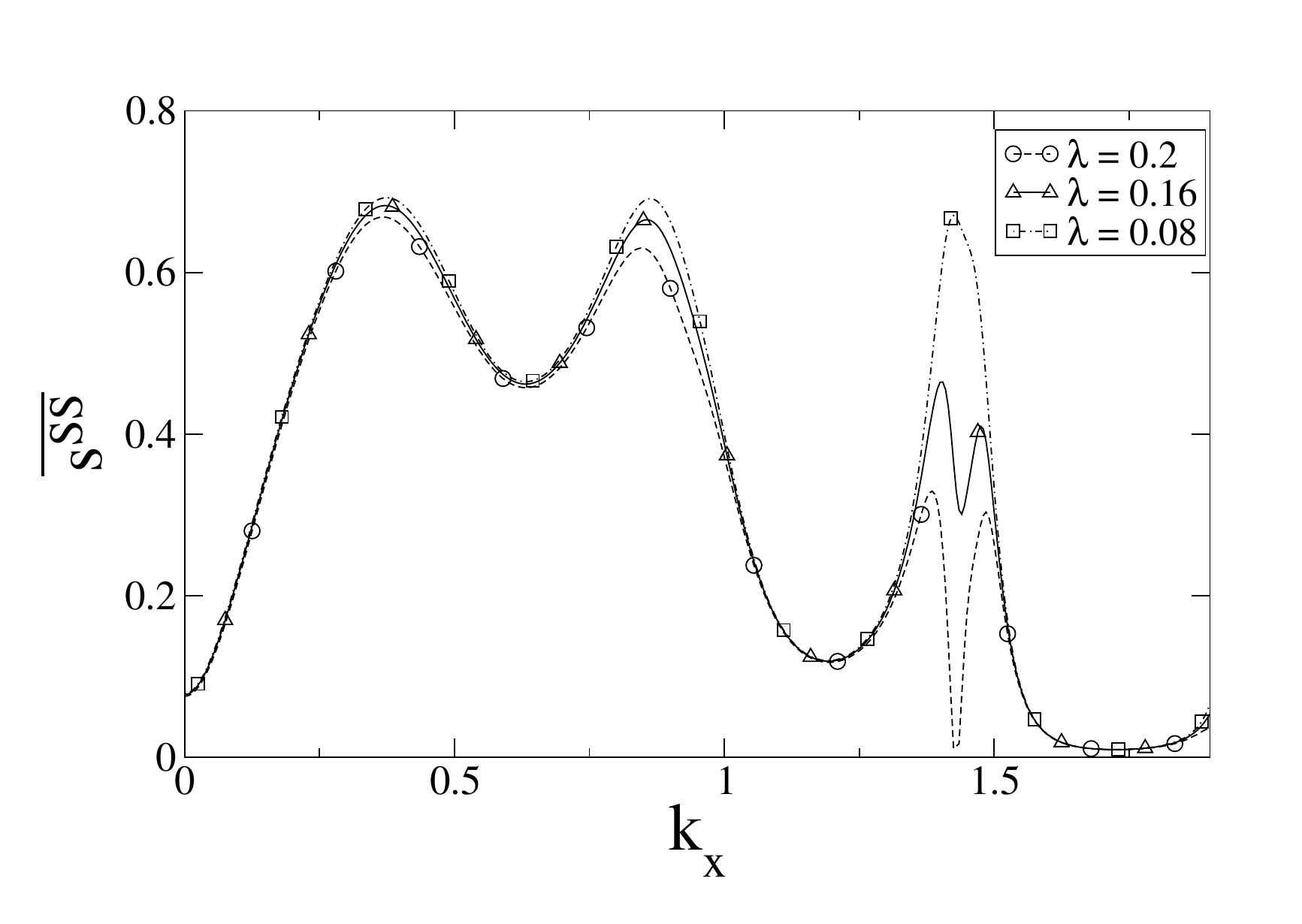}

\caption{Plot of the steady-state entropy after time-averaging, for three different electron-phonon
coupling strengths $\lambda/\Omega=0.08,0.16,0.2$ and phonon temperature $T=0.1\Omega$. We have set $\Omega=1.0$.
The entropy decreases with increasing strength of the coupling to the reservoir.}
\label{fig9}
\end{center}
\end{figure}

\begin{figure}
\begin{center}
\includegraphics[height=8cm,width=8cm,keepaspectratio]{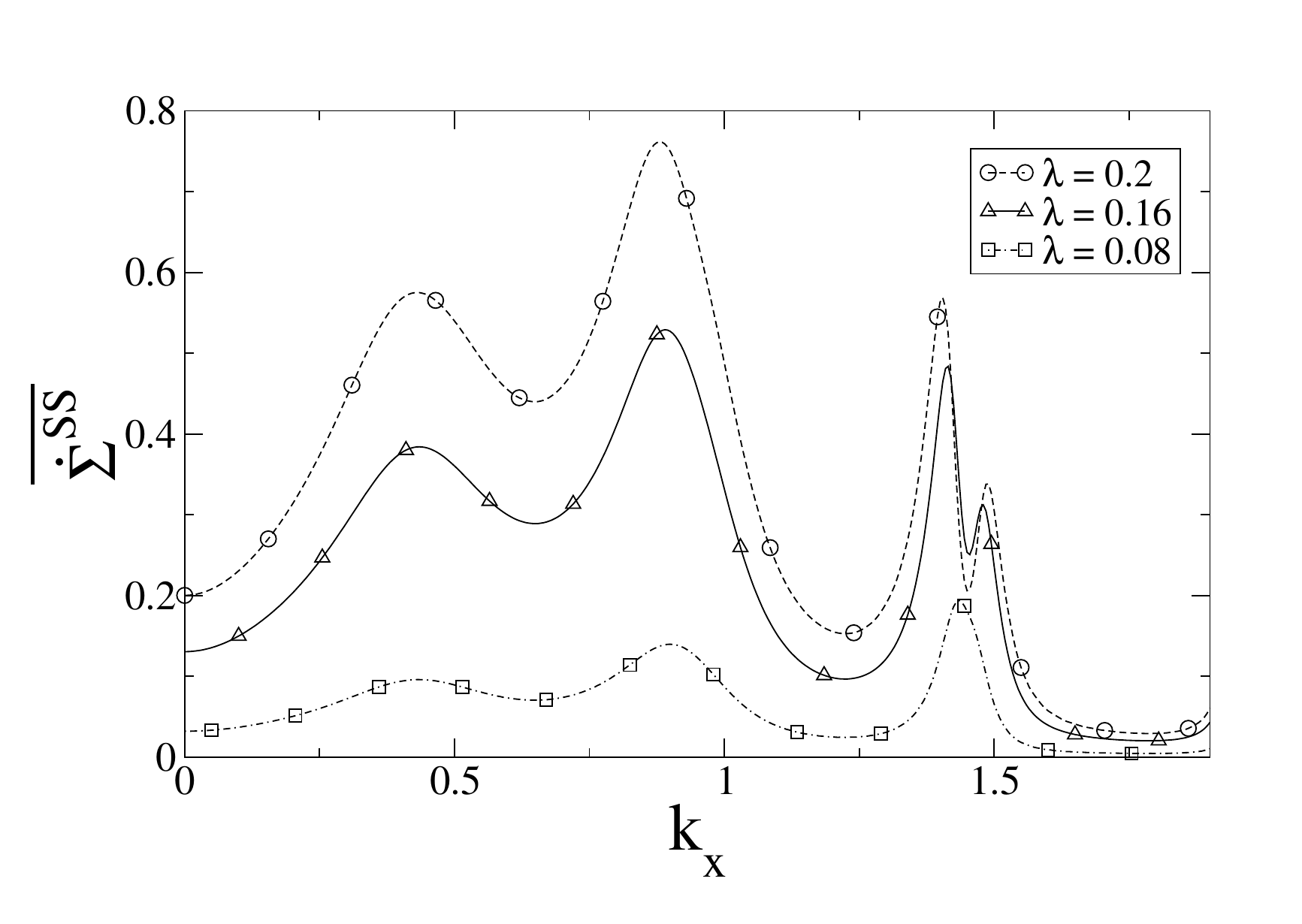}

\caption{The steady-state entropy production rate time-averaged over a laser cycle, along $k_y=0$ for three different
electron-phonon couplings $\lambda/\Omega=0.08,0.16,0.2$. The phonon temperature is $T=0.1\Omega$. We have set $\Omega=1.0$.
The entropy production rate increases as coupling to the reservoir is increased. This is accompanied by a decreasing decoherence (Fig.~\ref{fig8}), and
decreasing system entropy (Fig.~\ref{fig9}).}
\label{fig10}
\end{center}
\end{figure}

For non-zero $k$, where analytic computations are not possible anymore, our numerical results for the steady-state entropy and entropy production
rate are presented in Fig.~\ref{fig9} and Fig.~\ref{fig10}, respectively.
Fig.~\ref{fig9} shows that by increasing the coupling constant, the time-averaged entropy decreases. However, since the entropy is determined
mainly by the diagonal components of the density matrix which are not sensitive to the coupling constant,
the difference between different curves is small and is only significant around the resonances.
This decrease in the entropy can be interpreted effectively as a band gap opening induced by the electron-phonon coupling~\cite{Kohn09}
analogous to avoided level crossings caused by external perturbations.

Fig.~\ref{fig10} shows the time-averaged entropy production rate for different momenta and three different strengths of the coupling to the reservoir.
As the reservoir coupling increases, the steady-state entropy production rate increases indicating a larger
deviation from detailed balance. Recalling that the entropy production rate is proportional to the heat released by the system,
one naturally expects that by strengthening the coupling of the system to the reservoir, the outward heat generated by the system must increase.
Note that this figure has the same structure as Fig.~\ref{fig9} with an enhancement near resonances. However unlike Fig.~\ref{fig9}, in Fig.~\ref{fig10}
the curves are much more sensitive to the change in the coupling constant. This stems from the fact that, as proved in App.~\ref{appB},
the heat rate or entropy production rate depends mainly on the off-diagonal components of the density matrix, while the entropy depends mostly
on the diagonal components, where the latter are not very sensitive to the coupling constant. This also implies that a small
decrease in the steady-state entropy is compensated by a considerable amount of heat released by the system into the reservoir.

\section{Conclusions}\label{concl}

We have studied an open Floquet topological system under the assumption that the reservoir to which
the system is coupled is Markovian. The topology in the system arises because the circularly polarized laser opens
up a gap at the Dirac points whose origin is the breaking of time-reversal symmetry in the
Floquet Hamiltonian, with the quasi-energy bands acquiring a non-zero Berry curvature.

The combination of periodic drive and dissipation gives rise to many new
results. One of them is an effective reservoir induced coherence.
This comes about if there are reservoir density of states at the laser frequency, or
some multiples of it. For such a case, as the drive pumps energy into the system, the system can give up this energy to the reservoir.
In particular, this coherence arises due to Floquet-Umklapp processes that
allow the electron to release energy at multiples of the drive without causing transitions between distinct electronic states, where the
latter processes would be akin to decoherence.

The signature of this coherence is a non-zero off-diagonal matrix element in the Floquet basis that grows with the strength of the
coupling to the reservoir, while the diagonal component is only weakly affected by the coupling strength to the
reservoir. This has the effect of increasing
the purity of the steady-state density matrix, measured by 
${\rm Tr}\biggl[W_{\rm el}^2\biggr]$.

The second important result is a
steady-state that is in general not a Gibbs' distribution. We characterize this lack of detailed balance by a net steady-state entropy production rate.
We have shown that a non-zero entropy production mostly depends on the off-diagonal components of the density matrix.
Since in the steady-state an entropy production rate is equivalent to the heat released by the system, this result
can be used to engineer the phonon reservoir such that the system can efficiently give up heat to the reservoir, becoming
more coherent.
Furthermore we find that the entropy production rate increases with coupling to the reservoir, and also when one is closer to resonances,
where the quasi-energy level spacings become small.

Due to the synchronization of the system, we extract explicit results for the occupation probabilities of the Floquet levels
from a Fourier decomposition of the steady-state density  matrix. We show that in the vicinity of resonances,
not only do the off-diagonal elements become stronger, but more number of Floquet quasi-energy levels
are occupied.

All the above results are supplemented by exact analytic expressions at the Dirac point which highlight the complex interplay of
the many energy scales in the problem: system-reservoir coupling, reservoir temperature, frequency of the laser, amplitude of the laser,
and quasi-energies. Consequently one can use these different energy scales to find a criterion to maximize coherence.
The results for the Floquet occupation probabilities, and in particular their dependence on the strength of the
system-reservoir coupling, can be tested in experiments such as time-resolved and spin-resolved ARPES.

At this point, a note of caution is in order. We find that the Markovian approximation that relies on a weak-coupling to the reservoir,
breaks down for large momentum and for very small level crossings relative to
the coupling to the reservoir. For our parameters this happens near $k_x\sim 2$ and couplings greater than $\lambda/\Omega \sim 0.25$,
where the density matrix starts acquiring negative eigenvalues. In such a case,
more sophisticated methods which treat the reservoir non-perturbatively, are needed.

Floquet topological phase transitions are characterized by jumps in the Chern number. However the transitions when  measured
in terms of observables such as the Hall conductance will show rounding~\cite{Dehghani14}, with the extent of rounding depending
not only on the temperature of the reservoir but also on the strength of the coupling
to the reservoir. A proper theory for Floquet topological phase transitions for the open system will need to
account for the subtleties discussed in this paper.

One of the key observations of our paper is that if reservoir density of states exist at multiples of the drive frequency,
then very efficient cooling is possible, with the system at steady-state becoming more coherent than a Gibbs state at the temperature of the reservoir.
An interesting question would be to explore how the reservoir coupling can be further engineered to make the system reach dark states which
correspond to completely pure states with ${\rm Tr}\biggl[W_{\rm el}^2\biggr]=1$.

{\sl Acknowledgments:}
The authors thank A. Clerk, J. Keeling, S. Kehrein, I. Martin, and R. K. P. Zia  for helpful discussions.
This work was supported by the US Department of Energy,
Office of Science, Basic Energy Sciences, under Award No.~DE-SC0010821.

\appendix
\section{Synchronization of the steady-state density matrix} \label{appA}
We show in this appendix that after reaching the steady-state, the density matrix harmonizes with the external drive in the Schr\"odinger picture.
While in our numerical simulation we use a time dependent $R$ matrix, in this appendix we use a time-averaged rate matrix which will simplify the proof.
We start with the Floquet-Master Eq.~\eqref{rateo1gen} which consists of 4 complex equations for the components of the density matrix.
To represent these equations in a matrix form, we form a column vector from the components of the density matrix
\begin{eqnarray}
\boldsymbol{\vec{\rho}}_{k}^{D}(t) = \begin{pmatrix}\rho_{k,dd}(t)\\ \rho_{k,du}(t)\\\rho_{k,ud}(t)\\ \rho_{k,uu}(t)\end{pmatrix}.
\end{eqnarray}
Now the Floquet-Master equation can be represented in matrix form
\begin{eqnarray}
\boldsymbol{\dot{\vec{\rho}}}_{k}^{D}(t) = \boldsymbol{L}_{k}^{D}(t)\boldsymbol{\vec{\rho}}_{k}^{D}(t),\label{dependentRateEq}
\end{eqnarray}
where $\boldsymbol{L}_{k}^{D}(t)$ is a $4\times4$ matrix which can be read off from Eq.~\eqref{rateo1gen}.
Not all of the components of the above vector denoted in our notation by the superscript $D$ in $\boldsymbol{\vec{\rho}}_{k}^{D}$, are independent.
To find the linearly independent components of the density matrix note that the diagonal components of the density matrix are purely real and must satisfy
\begin{eqnarray}
\rho_{k,dd}+\rho_{k,uu} = 1.
\end{eqnarray}
To preserve this conservation equation, one of the eigenvalues of $\boldsymbol{L}_{k}^{D}(t)$ must be always equal to zero. Equivalently, we can remove one of the diagonal components of the density matrix e.g. $\rho_{k,uu} = 1 - \rho_{k,dd}$.
Moreover since the density matrix is Hermitian, this requires $\rho_{k,du} = \rho_{k,ud}^*$.
Thus we can define a linearly independent density vector consisting of 3 components as in the following
\begin{eqnarray}
\boldsymbol{\vec{\rho}}_{k}(t) = \begin{pmatrix}\rho_{k,dd}(t)\\ \rho_{k,du}(t)\\ \rho_{k,du}^*(t)\end{pmatrix}.
\end{eqnarray}
Eventually one can rewrite Eq.~\eqref{dependentRateEq} in terms of the independent components
\begin{eqnarray}
\boldsymbol{\dot{\vec{\rho}}}_{k}(t) = \boldsymbol{L}_{k}(t)\boldsymbol{\vec{\rho}}_{k}(t)+\boldsymbol{\vec{b}}_{k}(t).\label{rhoVecODE}
\end{eqnarray}
Above $\boldsymbol{L}_{k}(t)$ is a $3\times3$ time-dependent matrix which is formed by the rate matrix $R_{\alpha\beta, \gamma\delta}$
and the quasi-energy difference. We can separate these into two contributions,
\begin{eqnarray}
\boldsymbol{L}_{k}(t) = \boldsymbol{L}_{k}^{R}(t) + \boldsymbol{L}_{k}^{\epsilon},
\end{eqnarray}
where
\begin{eqnarray}
\boldsymbol{L}_{k}^{\epsilon} = \begin{pmatrix}0 & 0 & 0\\
0 & i\big(\epsilon_{u}-\epsilon_{d}\big) & 0\\
0 & 0 & i\big(\epsilon_{d}-\epsilon_{u}\big)
\end{pmatrix},
\end{eqnarray}
is purely imaginary and $\boldsymbol{L}_{k}^{R}(t)$ is only composed of $R_{\alpha\beta, \gamma\delta}^{k}$.
Here we do not need the explicit form of this matrix. In Eq.~\eqref{rhoVecODE}, $\boldsymbol{\vec{b}}_{k}(t)$ is a 3 component column vector which
originates from replacing $\rho_{k,uu}$ by $1-\rho_{k,dd}$.

Here we mention some of the properties of $\boldsymbol{L}^{R}_{k}(t)$ and $\boldsymbol{\vec{b}}_{k}(t)$ which will be used.
The first property is that since these two quantities are obtained from the components of the $R_{\alpha\beta, \gamma\delta}$ matrix, they can at most have a periodic
dependence on time. The second property has to do with the eigenvalues of the $\boldsymbol{L}_{k}(t)$ matrix. These can at most have real parts that are negative,
which ensure stable solutions where components of the density matrix are confined
($0\leq\rho_{k,dd}\leq1, |\rho_{k,du}| \leq 1/2$), and there is no exponential growth in time. This can be understood by considering the closed solutions of Eq.~\eqref{rhoVecODE}
\begin{eqnarray}
&&\boldsymbol{\vec{\rho}}_{k}(t) = e^{\int_{0}^{t}dt_{3}\boldsymbol{L}_{k}(t_{3})}\bigg(\boldsymbol{\vec{\rho}}_{k}(0)\nonumber\\
&&+\int_{0}^{t}dt_{1}e^{-\int_{0}^{t_{1}}dt_{2}\boldsymbol{L}_{k}(t_{2})}\boldsymbol{\vec{b}}_{k}(t_{1})\bigg).
\label{closedSol}
\end{eqnarray}

Now we plan to use the above information to study the asymptotic behavior of the solutions of the density matrix. However as one can see in the above formula,
without knowing the explicit time dependence of $\boldsymbol{L}_{k}(t)$, we cannot yet compute this integral.
To overcome this we consider the case where it is a permissible approximation to replace $\boldsymbol{L}_{k}(t)$
with its mean value. This requires that the temporal oscillations of $\boldsymbol{L}_{k}(t)$
around its mean value be comparatively small, which corresponds to small electron-phonon coupling constants and momentum.
In such cases one can verify by numerical simulations that the steady-state solutions of the original Floquet-Master equation
with a time-dependent rate matrix, after averaging over time yields the same answer as the solutions of the time-averaged
Floquet-Master equation with $\overline{\boldsymbol{L}}_{k}$. Therefore before considering the more general case we will consider a time-averaged master equation.

After replacing $\boldsymbol{L}_{k}(t)$ and $\boldsymbol{\vec{b}}_{k}(t)$ with their time-averaged values, we can calculate the time integrals explicitly
\begin{eqnarray}
&&\boldsymbol{\vec{\rho}}_{k}(t) = e^{t\overline{\boldsymbol{L}}_{k}}\bigg(\boldsymbol{\vec{\rho}}_{k}(0)+\int_{0}^{t}dt_{1}e^{-t_{1}\overline{\boldsymbol{L}}_{k}}\overline{\boldsymbol{\vec{b}}}_{k}\bigg).
\end{eqnarray}
In the steady-state, the first term in the above becomes infinitesimally small at long times because the eigenvalues of $\overline{\boldsymbol{L}_{k}}$
have negative real parts. Thus
\begin{eqnarray}
&&\lim_{t\rightarrow\infty}\boldsymbol{\vec{\rho}}_{k}(t) = \lim_{t\rightarrow\infty}e^{t\overline{\boldsymbol{L}}_{k}}\int_{0}^{t}dt_{1}e^{-t_{1}\overline{\boldsymbol{L}}_{k}}
\overline{\boldsymbol{\vec{b}}}_{k}.
\end{eqnarray}
We can compute the integrals and obtain
\begin{eqnarray}
\lim_{t\rightarrow\infty}\boldsymbol{\vec{\rho}}_{k}(t) =&&\lim_{t\rightarrow\infty} \overline{\boldsymbol{L}}_{k}^{-1}\Big(e^{t\overline{\boldsymbol{L}}_{k}}
- 1\Big)\overline{\boldsymbol{\vec{b}}}_{k}\notag\\
=&&-\overline{\boldsymbol{L}}_{k}^{-1}\overline{\boldsymbol{\vec{b}}}_{k}.
\end{eqnarray}
Therefore in the steady-state with a time-averaged $\overline{\boldsymbol{L}}_{k}$ the density matrix asymptotically becomes constant in time and the
oscillations with the frequency $\epsilon_{ku}-\epsilon_{kd}$ fade out.
Let us highlight that, this result does not depend on the specific form of the reservoir coupling.

After showing that the non-harmonic oscillations with $\epsilon_{ku} - \epsilon_{kd}$ vanish in a time-averaged Floquet-Master equation,
one can argue that in the time-dependent Floquet-Master equation such oscillations must vanish as well and one can only have harmonic solutions with frequency $\Omega$. This can be realized by using the Floquet theorem for differential equations. Very briefly this theorem states that the solution of a homogeneous linear differential equation with a periodic matrix, as in the time-dependent master equation Eq.~\eqref{dependentRateEq}, is given by $\boldsymbol{\vec{\rho}}_{k}^{D}(t) = \phi_{k}(t)\boldsymbol{\vec{\rho}}_{k}^{D}(0)$, where $\phi_{k}(t)$ is the fundamental solution of this differential equation which can be decomposed as $\phi_{k}(t)= P_{k}(t)e^{tB_{k}}$ with $P_{k}(t)$ a periodic in time matrix and $B_{k}$ a time-independent matrix~\cite{ODE1999}. As in $\boldsymbol{L}_{k}^{D}(t)$, one of the eigenvalues of $B_{k}$ must always vanish so as to preserve the conservation law of probabilities. Moreover, similar to the above discussion for the time-averaged $\overline{\boldsymbol{L}}_{k}$, one can argue that the other three eivenvalues of $B_{k}$ must have negative real parts. Therefore at long times, $e^{tB_{k}}$ will asymptotically become constant, and therefore only the periodic part of the fundamental solution will survive~\cite{Kohn09}. This proves our claim for
a Floquet-Master equation with periodic in time rates.

\section{Steady-state entropy production rate} \label{appB}
Here we plan to derive an expression for the steady-state entropy production rate, namely Eq.~\eqref{entRateSSFinalm}.
Let us start from the following expression derived in the main text,
\begin{equation}
\dot{\Sigma}=-{\rm Tr}\biggl[\dot{W}_{\rm el}\ln\biggl(\dfrac{W_{\rm el}}{W_{\rm el}^{\rm eq}}\biggl)\biggl],\label{eq:EntProd}
\end{equation}
where
\begin{equation}
W^{\rm eq}_{\rm el}=\dfrac{e^{-\beta H_{\rm el}(t)}}{Z_{t}},\quad Z_{t}=\mbox{Tr}\bigg[e^{-\beta H_{\rm el}(t)}\bigg],
\end{equation}
$H_{\rm el}(t)$ denotes the time dependent Hamiltonian of the system.
We can simplify Eq.~\eqref{eq:EntProd} by expanding the logarithm
\begin{eqnarray}
&&\dot{\Sigma} =
-\mbox{Tr}\Big[\dot{W}_{\rm el}\ln W_{\rm el}\Big]-\mbox{Tr}\Big[\beta H_{\rm el}(t)\dot{W}_{\rm el}\Big]\nonumber\\
&&-\Big(\ln Z_{t}\Big)\mbox{Tr}\Big[\dot{W}_{\rm el}\Big].\label{sigmaExpanded}
\end{eqnarray}
We are only interested in the average of the entropy production rate at steady-state where,
the density matrix synchronizes with the external drive, and the density matrix in the
Schr\"odinger picture is
\begin{equation}
W_{\rm el}^{\rm SS}(t)=\sum_{\alpha,\beta}\rho_{k,\alpha\beta}^{\rm SS}|\phi_{k,\alpha}(t)\rangle\langle\phi_{k,\beta}(t)|.
\end{equation}
As we have shown in Appendix~\ref{appA}, in the Schr\"odinger picture all the components of $\rho_{k, \alpha\beta}^{\rm SS}$,
including the off-diagonal components,
are periodic in the steady-state. As the Floquet quasi-modes are periodic too, one finds $W_{\rm el}^{\rm SS}(t+T_{\Omega}) = W_{\rm el}^{\rm SS}(t)$.
Consequently by periodicity of $W_{\rm el}^{\rm SS}$ we find that on time-averaging Eq.~\eqref{sigmaExpanded} over a cycle of the laser,
\begin{eqnarray}
\overline{\dot{\Sigma}^{\rm SS}}\equiv \frac{1}{T_{\Omega}}\int_{0}^{T_{\Omega}}\dot{\Sigma}^{\rm SS}dt
= -\beta{\rm Tr}\biggl[\overline{H_{\rm el}(t)\dot{W}_{\rm el}^{\rm SS}(t)}\biggl]
\label{entRateSS}.
\end{eqnarray}
The time derivative of the density matrix in the Schr\"odinger picture is
\begin{eqnarray}
&&\dot{W}_{\rm el}^{\rm SS}= \sum_{\alpha,\beta}\bigg[\dot{\rho}_{k, \alpha\beta}^{\rm SS}|\phi_{k,\alpha}(t)\rangle\langle\phi_{k,\beta}(t)|+\notag\\&&\rho_{k, \alpha\beta}^{\rm SS}|\dot{\phi}{}_{k,\alpha}(t)\rangle\langle\phi_{k,\beta}(t)|+\rho_{k, \alpha\beta}^{\rm SS}|\phi_{k,\alpha}(t)\rangle\langle\dot{\phi}{}_{k,\beta}(t)|\bigg].
\end{eqnarray}
To proceed we must compute the trace in
Eq.~\eqref{entRateSS}. Note that with an arbitrary operator $A$, and arbitrary vectors $u$ and $v$, the definition of tracing gives
\begin{equation}
{\rm Tr}\biggl[H_{\rm el}(t)|u\rangle \langle v|\biggl] = \langle v|H_{\rm el}(t)|u\rangle.
\label{entRateSS1}
\end{equation}
This simplifies Eq.~\eqref{entRateSS}
\begin{eqnarray}
\overline{\dot{\Sigma}^{\rm SS}} &&= -\beta\sum_{\alpha,\beta}\frac{1}{T_{\Omega}}\int_{0}^{T_{\Omega}}dt\bigg[\dot{\rho}_{k, \alpha\beta}^{\rm SS}\langle\phi_{k,\beta}(t)|H_{\rm el}(t)|\phi_{k,\alpha}(t)\rangle\notag\\&&+\rho_{k, \alpha\beta}^{\rm SS}\langle\phi_{k,\beta}(t)|H_{\rm el}(t)|\dot{\phi}{}_{k,\alpha}(t)\rangle\nonumber\\
&&+\rho_{k, \alpha\beta}^{\rm SS}\langle\dot{\phi}{}_{k,\beta}(t)|H_{\rm el}(t)|\phi_{k,\alpha}(t)\rangle\bigg].
\label{entRateSSTrace}
\end{eqnarray}
The action of the Hamiltonian on Floquet states can be computed by using the definition of the Floquet Hamiltonian
\begin{eqnarray}
H^{F}_{\rm el}=H_{\rm el}(t) - i\partial_{t},
\end{eqnarray}
More explicitly, this gives
\begin{eqnarray}
H_{\rm el}(t)|\phi_{k,\alpha}(t)\rangle=&&H^{F}_{\rm el}|\phi_{k,\alpha}(t)\rangle+ i\partial_{t}|\phi_{k,\alpha}(t)\rangle\notag\\=&&
\epsilon_{k\alpha}|\phi_{k,\alpha}(t)\rangle+ i|\dot{\phi}_{k,\alpha}(t)\rangle,
\end{eqnarray}
and its complex conjugate
\begin{eqnarray}
\langle\phi_{k,\alpha}(t)|H_{\rm el}(t)=&&\langle\phi_{k,\alpha}(t)|\epsilon_{k\alpha}- i\langle\dot{\phi}_{k,\alpha}(t)|.
\end{eqnarray}
We can insert these relations in Eq.~\eqref{entRateSSTrace}. Let us consider each term in this equation separately. The first term becomes
\begin{eqnarray}
\overline{\dot{\Sigma}^{\rm SS}_{1}} &&\equiv -\beta\sum_{\alpha,\beta}\frac{1}{T_{\Omega}}\int_{0}^{T_{\Omega}}dt\dot{\rho}_{k, \alpha\beta}^{\rm SS}\langle\phi_{k,\beta}(t)|H_{\rm el}(t)|\phi_{k,\alpha}(t)\rangle\notag.\\
&& = -\beta\sum_{\alpha,\beta}\Bigr[\epsilon_{k\alpha}\delta_{\alpha\beta}\overline{\dot{\rho}_{k, \alpha\beta}^{\rm SS}}
+i\overline{\dot{\rho}_{k, \alpha\beta}^{\rm SS}\langle\phi_{k,\beta}(t)|\dot{\phi}_{k,\alpha}(t)\rangle}\Bigl]\notag\\
&& = -\beta\sum_{\alpha,\beta}i\overline{\dot{\rho}_{k, \alpha\beta}^{\rm SS}\langle\phi_{k,\beta}(t)|\dot{\phi}_{k,\alpha}(t)\rangle},
\label{entRateSSTrace1}
\end{eqnarray}
where in the last equality we have used that $\overline{\dot{\rho}_{k, \alpha\beta}^{\rm SS}}$ vanishes
because in the steady-state the density matrix is periodic, and its time-averaged value is constant. Thus the time derivative of the average vanishes.
The second and third terms in Eq.~\eqref{entRateSSTrace} are respectively given by
\begin{eqnarray}
&&\overline{\dot{\Sigma}^{\rm SS}_{2}} \equiv -\beta\sum_{\alpha,\beta}\frac{1}{T_{\Omega}}\int_{0}^{T_{\Omega}}dt\rho_{k, \alpha\beta}^{\rm SS}\langle\phi_{k,\beta}(t)|H_{\rm el}(t)|\dot{\phi}_{k,\alpha}(t)\rangle\notag\\
&&= -\beta\sum_{\alpha,\beta}\frac{1}{T_{\Omega}}\int_{0}^{T_{\Omega}}dt\times\notag\\
&&\Bigr[\epsilon_{k\beta}\rho_{k, \alpha\beta}^{\rm SS}\langle\phi_{k,\beta}(t)|\dot{\phi}_{k,\alpha}(t)\rangle-i\rho_{k, \alpha\beta}^{\rm SS}\langle\dot{\phi}_{k,\beta}(t)|\dot{\phi}_{k,\alpha}(t)\rangle\Bigl].\notag\\
\label{entRateSSTrace2}
\end{eqnarray}
and
\begin{eqnarray}
&&\overline{\dot{\Sigma}^{\rm SS}_{3}} \equiv -\beta\sum_{\alpha,\beta}\frac{1}{T_{\Omega}}\int_{0}^{T_{\Omega}}dt\rho_{k, \alpha\beta}^{\rm SS}\langle\dot{\phi}_{k,\beta}(t)|H_{\rm el}(t)|\phi_{k,\alpha}(t)\rangle\notag\\
&&= -\beta\sum_{\alpha,\beta}\frac{1}{T_{\Omega}}\int_{0}^{T_{\Omega}}dt\times\notag\\
&&\Bigr[\epsilon_{k\alpha}\rho_{k, \alpha\beta}^{\rm SS}\langle\dot{\phi}_{k,\beta}(t)|\phi_{k,\alpha}(t)\rangle+i\rho_{k, \alpha\beta}^{\rm SS}\langle\dot{\phi}_{k,\beta}(t)|\dot{\phi}_{k,\alpha}(t)\rangle\Bigl].\notag\\
\label{entRateSSTrace3}
\end{eqnarray}
By summing the last two equations, terms with opposite signs cancel, and we obtain
\begin{eqnarray}
&&\overline{\dot{\Sigma}^{\rm SS}_{2}}+\overline{\dot{\Sigma}^{\rm SS}_{3}}= -\beta\sum_{\alpha,\beta}\bigl(\epsilon_{k\beta}-\epsilon_{k\alpha}\bigr)\overline{\rho_{k, \alpha\beta}^{\rm SS}\langle\phi_{k,\beta}(t)|\dot{\phi}_{k,\alpha}(t)\rangle},\notag\\
\label{entRateSSTrace2sum3}
\end{eqnarray}
where in the last equation we have used that $\partial_{t}\langle\phi_{k,\alpha}(t)|\phi_{k,\beta}(t)\rangle=0$, so that,
\begin{equation}
\langle\dot{\phi}_{k,\alpha}(t)|\phi_{k,\beta}(t)\rangle=-\langle\phi_{k,\alpha}(t)|\dot{\phi}{}_{k,\beta}(t)\rangle.
\end{equation}
Compiling the results from Eq.~\eqref{entRateSSTrace2sum3} and Eq.~\eqref{entRateSSTrace1}, we find
\begin{eqnarray}
&&\overline{\dot{\Sigma}^{\rm SS}}= -\beta\sum_{\alpha,\beta}\overline{\langle\phi_{k,\beta}(t)|\dot{\phi}_{k,\alpha}(t)\rangle\bigl(\epsilon_{k\beta}-\epsilon_{k\alpha}+i\partial_{t}\bigr)\rho_{k, \alpha\beta}^{\rm SS}}.\notag\\
\label{entRateSSFinal}
\end{eqnarray}
It is straightforward to check that the right hand side of Eq.~\eqref{entRateSSFinal} is a purely real quantity.

Before using this result for a two-level system, we must explain some of its properties. First, note that since the entropy production rate is a
physical quantity, the left hand side of this equation must be gauge invariant. Recall that while the Floquet quasi-modes and quasi-energies are not unique,
the Schr\"odinger wave functions which are given by
\begin{eqnarray}
|\psi_{k\alpha}(t)\rangle = e^{-i\epsilon_{k\alpha}t}|\phi_{k\alpha}\rangle.
\end{eqnarray}
are invariant under the following gauge transformations
\begin{eqnarray}
|\phi_{k\alpha}(t)\rangle &&\rightarrow e^{i m_{\alpha}\Omega t}|\phi_{k\alpha}(t)\rangle,\\
\epsilon_{k\alpha} &&\rightarrow \epsilon_{k\alpha} + m_{\alpha}\Omega,
\end{eqnarray}
where $m_{\alpha}$ is an integer. Any physical observable is obtained from taking a trace with the density matrix $W_{\rm el}$. Since the
result should be gauge invariant, this requires that under the above transformations, $\rho_{\alpha\beta}$ must transform as
\begin{eqnarray}
\rho_{\alpha\beta} &&\rightarrow e^{-i (m_{\alpha}-m_{\beta})\Omega t}\rho_{\alpha\beta}.
\end{eqnarray}
By applying the above consideration to Eq.~\eqref{entRateSSFinal}, one can easily check that this equation satisfies the necessary condition of
gauge independence. More importantly, this result shows that the oscillating part of the off-diagonal component can be as
important as its time-averaged value. This also implies that by using a time-averaged Floquet-Master equation, where the oscillations are ignored,
some important information about physical quantities could be neglected.

Now we consider the case of a two-level system.
Separating the contribution of diagonal and off-diagonal density matrix components,
\begin{eqnarray}
&&\overline{\dot{\Sigma}^{\rm SS}}= -\beta\Bigr[\overline{\sum_{\alpha\neq\bar\alpha}\langle\phi_{k,\bar\alpha}(t)|\dot{\phi}_{k,\alpha}(t)\rangle
\bigl(\epsilon_{k\bar\alpha}-\epsilon_{k\alpha}+i\partial_t\bigr)
\rho_{k, \alpha\bar\alpha}^{\rm SS}}\notag\\
&&+\sum_{\alpha}i\overline{\dot{\rho}_{k, \alpha\alpha}^{\rm SS}\langle\phi_{k,\alpha}(t)|\dot{\phi}_{k,\alpha}(t)\rangle}\Bigl],
\label{entRateSSSplit}
\end{eqnarray}
and writing the above in terms of $d$ and $u$ states, we obtain,
\begin{eqnarray}
&&\overline{\dot{\Sigma}^{\rm SS}} = 2\beta\nonumber\\
&&\times \mbox{Re}\Big[\overline{\left(\left(\epsilon_{ku} - \epsilon_{kd}\right)\rho_{k,du}^{\rm SS} + i\dot{\rho}_{k,du}^{\rm SS}\right)\langle\dot{\phi}{}_{k,u}(t)|\phi_{k,d}(t)\rangle}\Big]
\notag\\
&&-\beta\mbox{Im}\Bigr[\overline{\dot{\rho}_{k,dd}^{\rm SS}\langle\dot{\phi}_{k,d}(t)|\phi_{k,d}(t)\rangle}-\overline{\dot{\rho}_{k,dd}^{\rm SS}\langle\dot{\phi}_{k,u}(t)|\phi_{k,u}(t)\rangle}\Bigl],\notag\\
\label{TwoLevelDtSigma}
\end{eqnarray}
where in the last line we have used that $\dot{\rho}_{k,uu}^{\rm SS} = -\dot{\rho}_{k,dd}^{\rm SS}$.

Note that in our simulations, $\dot{\rho}_{k,dd}^{\rm SS}$ which can be read from the Fourier expansions $\rho_{k,dd}^{n \rm SS}$ in Fig.~\ref{fig4},~\ref{fig5},
is almost negligible compared to the off-diagonal component. As a consequence the main contribution of the entropy production rate originates from the
off-diagonal component of the density matrix. We can also see this at $k=0$ where exact analytic expressions exist.
In particular for weak couplings (as compared to $\Omega$), where the steady-state diagonal density matrix elements have a weak oscillation
amplitude, the entropy production can be approximated by
\begin{eqnarray}
&&\overline{\dot{\Sigma}^{\rm SS}} \approx 2\beta\nonumber\\
&&\times \mbox{Re}\Big[\overline{\left(\left(\epsilon_{ku} - \epsilon_{kd}\right)\rho_{k,du}^{\rm SS} + i\dot{\rho}_{k,du}^{\rm SS}\right)\langle\dot{\phi}{}_{k,u}(t)|\phi_{k,d}(t)\rangle}\Big].\notag\\
\label{TwoLevelDtSigmaApprox}
\end{eqnarray}

\section{Analytic results near the Dirac point ($\boldsymbol{k=0}$)} \label{appC}
In this section we give some intermediate steps in the derivation of the analytic solutions near the Dirac point.
At $k=0$, the Hamiltonian simplifies to,
\begin{eqnarray}
H_{\rm el}(k=0,t)= A_0\begin{pmatrix}0 & e^{i\Omega t}\\ e^{-i\Omega t}&0\end{pmatrix},
\end{eqnarray}
and exact expressions can be obtained for the Floquet modes.~\cite{Dehghani14} The
density matrix at $k=0$ is (in this sub-section we will suppress the $k$ label),
\begin{eqnarray}
W_{\rm el}=\sum_{\alpha\beta}\rho_{\alpha\beta}^{I}e^{-i\left(\epsilon_{\alpha}-\epsilon_{\beta}\right)t}|\phi_{\alpha}(t)\rangle\langle
\phi_{\beta}(t)|.
\end{eqnarray}
Note that there are a multiplicity of quasi-energy levels, yet there are only two distinct exact eigenstates of the periodic Hamiltonian
that we label by "up (u)" and "down (d)" in the main text. The quasi-modes $|\phi_{u,d}(t)\rangle$ and the exact eigenstates
$|\psi_{u,d}(t)\rangle$ are related as $|\psi_{u,d}(t)\rangle = e^{-i\epsilon_{u,d} t} |\phi_{u,d}(t)\rangle$. As briefly shown in Appendix~\ref{appB},
 Floquet quasi-modes and quasi-energies are not uniquely determined. For $k=0$, the quasi-energies may be written as,
\begin{eqnarray}
&&\epsilon_{d}=m_{d}\Omega+\frac{-\Omega-\Delta}{2},\quad\epsilon_{u}=m_{u}\Omega+\frac{-\Omega+\Delta}{2},
\end{eqnarray}
where $\Delta =\sqrt{4 A_0^2 + \Omega^2}$ and $m_{d,u}$ are arbitrary integers. The corresponding quasi-modes are then given by
\begin{eqnarray}
&&|\phi_{d}(t)\rangle=e^{im_{d}\Omega t}\begin{pmatrix}d_{1u}\\ e^{-i\Omega t}d_{2u}\end{pmatrix},\nonumber \\
&&|\phi_{u}(t)\rangle=e^{im_{u}\Omega t}\begin{pmatrix}d_{1d}\\ e^{-i\Omega t}d_{2d}\end{pmatrix},\nonumber\\
\label{FloqStates}
\end{eqnarray}
where
\begin{eqnarray}
&&d_{1u}= \frac{\sqrt{2}A_0}{\sqrt{\Delta \left(\Delta - \Omega\right)}}\,\,;d_{2u}= \frac{1}{\sqrt{2}}\sqrt{1- \frac{\Omega}{\Delta}},\\
&&d_{1d}= \frac{\sqrt{2}A_0}{\sqrt{\Delta \left(\Delta + \Omega\right)}}\,\,;d_{2d}= -\frac{1}{\sqrt{2}}\sqrt{1+ \frac{\Omega}{\Delta}}.
\end{eqnarray}
Later we will choose a gauge in which $m_{d} = m_{u} = 1$. This  results in $\epsilon_{u} - \epsilon_{d} = \Delta > \Omega$. The reason for choosing this seemingly
unnatural gauge is that as will be clarified later, in this gauge, the transition rates are time-independent.
However, to avoid the problem of gauge-dependence, it is often convenient to construct the matrix
elements not between the quasi-modes $|\phi_{u,d}(t)\rangle$ themselves as done in the main text, but between the exact eigenstates $|\psi_{u,d}(t)\rangle$.
We refer to the matrix elements between the exact eigenstates as the gauge-invariant matrix elements, and the entire Floquet-Master
equation can be written in terms of them.
The gauge-invariant matrix elements for our model are $C^{\rm gi}_{1,2\alpha\beta}=C_{1,2 \alpha \beta}e^{i(\epsilon_{\alpha}-\epsilon_{\beta})t}$,
with the $C_{1,2\alpha\beta}$
being the matrix elements between the
quasi-modes. We may write,
\begin{eqnarray}
&&C_{1\alpha\beta}^{\rm gi}(t) = e^{i\left(\epsilon_{\alpha}-\epsilon_{\beta}\right)t}\langle \phi_{\alpha}(t)|c_{\uparrow}^{\dagger}c_{\downarrow}|\phi_{\beta}(t)\rangle,\\
&&C_{2\alpha\beta}^{\rm gi}(t) = e^{i\left(\epsilon_{\alpha}-\epsilon_{\beta}\right)t} \langle \phi_{\alpha}(t)|c_{\downarrow}^{\dagger}c_{\uparrow}|\phi_{\beta}(t)\rangle.
\end{eqnarray}
At the Dirac point we find,
\begin{eqnarray}
C_{1uu}(t)=\frac{A_0}{\Delta}e^{-i \Omega t},\\
C_{1dd}(t)=-\frac{A_0}{\Delta}e^{-i\Omega t},\\
C_{2uu}(t)=\frac{A_0}{\Delta}e^{i\Omega t},\\
C_{2dd}(t)=-\frac{A_0}{\Delta}e^{i\Omega t},\\
C_{1ud}^{\rm gi}(t)= -\frac{1}{2}\left(1+\frac{\Omega}{\Delta}\right)e^{-i\Omega t+i\Delta t},\\
C_{1du}^{\rm gi}(t)=\frac{1}{2}\left(1-\frac{\Omega}{\Delta}\right)
e^{-i\Omega t -i\Delta t},\\
C_{2ud}^{\rm gi}(t)=\frac{1}{2}\left(1-\frac{\Omega}{\Delta}\right)
e^{i\Omega t + i \Delta t}, \\
C_{2du}^{\rm gi}(t)=-\frac{1}{2}\left(1+\frac{\Omega}{\Delta}\right)
e^{i\Omega t-i\Delta t}.
\end{eqnarray}
It is convenient to define the corresponding gauge-invariant rates,
\begin{eqnarray}
R_{ab,cd}^{\rm gi}= e^{it(\epsilon_a-\epsilon_b+\epsilon_c-\epsilon_d)}R_{ab,cd}.
\end{eqnarray}
The rate equation~\eqref{rateo1gen}, can be recast in terms of these gauge-invariant rates as follows:
\begin{eqnarray}
&&\dot{\rho}_{k,\alpha\beta}^{I}(t)=-\sum_{\delta \gamma}\biggl[R_{\alpha \delta,\delta \gamma}^{\rm gi}(t)\rho_{k, \gamma \beta}^{I}(t)+\ldots\biggr],
\end{eqnarray}
where $\rho_{k, \alpha\beta}^{S}=\rho_{k, \alpha\beta}^{I}e^{-it(\epsilon_{\alpha}-\epsilon_{\beta})}$.
In the following we will solve the Floquet-Master equation for the density matrix in the Schr\"odinger picture, $\rho_{k, \alpha\beta}^{S}$.

Now one can use the freedom in choosing the quasi-energy levels and the
corresponding quasi-modes, such that the rates $R$ become time-independent. This is performed by choosing $|\epsilon_{\alpha}-\epsilon_{\beta}|=\Delta\left(1-\delta_{\alpha\beta}\right)$, because only $n_1=n_2=\pm 1$ terms survive in Eq.~\eqref{Rdef}.
Moreover, in the limit of $A_0/\Omega\ll 1$, where $1-\frac{\Omega}{\Delta}\simeq \frac{2 A_0^2}{\Omega^2},
1+\frac{\Omega}{\Delta}\simeq 2 + {\cal O}
\left(\frac{A_0^2}{\Omega^2}\right)$, we find the following expressions for the rates,
\begin{widetext}
\begin{eqnarray}
&&R_{uu,uu}= 2\lambda^2\nu \frac{A_0^2}{\Omega^2}\biggl(1+2N_0\biggr)= R_{dd,dd} =-R_{uu,dd}=-R_{dd,uu},\\
&&R_{uu,ud}=  -2\lambda^2\nu \frac{A_0}{\Omega}\biggl[N_--\frac{A_0^2}{\Omega^2}N_+\biggr]\simeq -2\lambda^2\nu \frac{A_0}{\Omega}N_-;\,\,
R_{uu,du}= -2\lambda^2\nu \frac{A_0}{\Omega}\biggl(1+N_-\biggr),\\
&&R_{ud,uu}= -2\lambda^2\nu \frac{A_0}{\Omega}\biggl[N_0-\frac{A_0^2}{\Omega^2}\left(1+N_0\right)\biggr];\,\,
R_{du,uu}= -2\lambda^2\nu \frac{A_0}{\Omega}\biggl[\left(1+N_0\right)- \frac{A_0^2}{\Omega^2}N_0 \biggr]\simeq -2\lambda^2\nu \frac{A_0}{\Omega}\left(1+N_0\right),\\
&&R_{du,ud}= 2\lambda^2\nu N_-;\,\,
R_{ud,du}= 2\lambda^2\nu \biggl(1+N_-\biggr);\,\,
R_{ud,ud}= 2\lambda^2\nu \frac{A_0^2}{\Omega^2}\biggl(N_-+N_+\biggr),\\
&&R_{du,du}= -2\lambda^2\nu \frac{A_0^2}{\Omega^2}\biggl(2+ N_-+N_+\biggr);\,\,
R_{ud,dd}=2\lambda^2\nu \frac{A_0}{\Omega}\biggl[N_0-\frac{A_0^2}{\Omega^2}\left(1+N_0\right)\biggr],\\
&&R_{dd,du}= 2\lambda^2\nu \frac{A_0}{\Omega}\biggl(1+ N_-\biggr);\,\,
R_{du,dd}=-2\lambda^2\nu\frac{A_0}{\Omega}\biggl[\frac{A_0^2}{\Omega^2}N_0-(1+N_0)\biggr]\simeq 2\lambda^2\nu\frac{A_0}{\Omega}\left(1+N_0\right),\\
&&R_{dd,ud}=-2\lambda^2\nu\frac{A_0}{\Omega}\biggl[\frac{A_0^2}{\Omega^2}N_+-N_-\biggr]\simeq 2\lambda^2\nu\frac{A_0}{\Omega}N_-.
\end{eqnarray}
\end{widetext}
Above $N_0= N\left(\Omega\right),N_{\pm}=N\left(\Delta\pm \Omega\right)$.

In general in the steady-state, the density matrix can oscillate with frequency $\Omega$. However for $k=0$, as one sees above, in this gauge
all the scattering rates are constant in time as they do not contain oscillating terms. Therefore from the discussion of Appendix~\ref{appA},
one can deduce that at $k=0$, the steady-state density matrix attains a constant value $\partial_{t}\rho_{k=0,\alpha\beta}^{\rm SS} = 0$.
Using this in Eq.~\eqref{rateo1gen}, one finds, that the steady-state diagonal and off-diagonal components of the
density matrix in the Schr\"odinger picture are related as follows
\begin{eqnarray}
&&\rho_{dd}^{\rm SS}= \frac{R_{ud,du}+ {\rm Re}\left[\left(R_{ud,dd}-R_{du,uu}\right) \rho_{du}^{\rm SS}\right]}{R_{ud,du}+ R_{du,ud}},\\
&&0=i\left(\epsilon_u-\epsilon_d\right)\rho_{du}^{\rm SS} + \left(R_{ud,uu}+2 R_{uu,du}-R_{du,uu}\right)\nonumber\\
&&+ \rho_{du}^{\rm SS}\left(R_{ud,ud}+R_{du,du}\right)+\rho_{du}^{\rm SS}\left(2R_{dd,uu}-2R_{uu,uu}\right.\nonumber\\
&&\left.-R_{ud,du} -R_{du,ud}\right)+\rho_{dd}^{\rm SS}\left(2R_{dd,ud}-2R_{uu,du}\right).
\end{eqnarray}
The above equations may be used to solve for all the components of the steady-state reduced density matrix.
Here we simply note that at temperatures
small as compared to the quasi-energy level spacing $|\epsilon_u-\epsilon_d|$,
\begin{eqnarray}
\rho_{dd}^{\rm SS} = 1 + {\cal O} \biggl(\frac{A_0}{\Omega}{\rm Re}\left[\rho_{du}^{\rm SS}\right]\biggr),
\end{eqnarray}
where
\begin{eqnarray}
&&{\rm Re}\left[\rho_{du}^{\rm SS}\right]={\cal O}\biggl(\frac{\lambda^2\nu }{|\epsilon_d-\epsilon_u|}{\rm Im}\bigg[\rho_{du}^{\rm SS}\biggr]\biggr).
\end{eqnarray}

Below we give explicit results only
for the imaginary part of the off-diagonal component, because as we show below, it is only this component that
enters in the steady-state entropy production rate at $k=0$. After some algebra, we find that for $A_{0}/\Omega \ll 1$,
\begin{eqnarray}
{\rm Im}\bigl[\rho_{du}^{\rm SS}\bigr] \approx\ -\frac{\big(\epsilon_u-\epsilon_d\big)\big(R_{ud,dd}+R_{du,uu}\big)}{\big(\epsilon_u-\epsilon_d\big)^2+\big(R_{du,ud}+R_{ud,du}\big)^2}\notag.\\\label{ImRho}
\end{eqnarray}
By looking at the $R$ matrix components in the numerator, one can determine the processes which play a significant role in the entropy production.
From Eq.~\eqref{Rdef} for $R_{\alpha\beta,\alpha'\beta'}^{k}$, the energy conservation requires that the change in the energy of
the electrons by $\epsilon_{k\beta'}-\epsilon_{k\alpha'}- n\Omega$ must be supplied by the reservoir.
Here since we have $R_{ud,dd}+R_{du,uu}$ in the numerator, the immediate conclusion is that, here we have Floquet-Umklapp processes where the initial and final states of
the electrons are the same, and correspond to absorbed or emitted phonons with an energy equal to some multiple of the laser frequency.
After rewriting Eq.~\eqref{ImRho} explicitly in terms of the amplitude and frequency of the drive we find
\begin{eqnarray}
{\rm Im}\bigl[\rho_{du}^{\rm SS}\bigr] =2\lambda_{\Omega}^2\nu_{\Omega} \frac{A_0}{\Omega^2 + \left(2\lambda_{-}^2\nu_{-}\right)^2\left(1+2N_-\right)^2},\label{offd3}
\end{eqnarray}
where the subscripts ${\Omega}$ and $-$ in $\lambda$ and $\nu$ denote the value of these quantities, and hence the reservoir density of states
at energy $\Omega$ and energy $\Omega_{-} = \Delta - \Omega \approx 2A_{0}^2/\Omega$, respectively. Note that $\Omega_-$ is
the topological gap at the Dirac point.

From above it is clear that at low temperatures compared to the topological gap, so that the Bose function is small, the real part
of the off-diagonal density matrix is
\begin{eqnarray}
&&{\rm Re}\left[\rho_{du}^{\rm SS}\right]={\cal O}\biggl(\frac{\lambda^4\nu^2 A_{0}}{\Omega^{3}}\biggr).
\end{eqnarray}
The entropy production rate as derived in Eq.~\eqref{TwoLevelDtSigma}, for a constant in time steady-state density matrix is
\begin{eqnarray}
&&\overline{\dot{\Sigma}^{\rm SS}} = 2\beta\left(\epsilon_{u} - \epsilon_{d}\right)\mbox{Re}\bigg[\overline{\rho_{du}^{\rm SS}\langle\dot{\phi}_{k,u}(t)|\phi_{k,d}(t)\rangle}\bigg].
\end{eqnarray}
From Eq.~\eqref{FloqStates}, one finds
\begin{eqnarray}
\langle\dot{\phi}_{k,u}(t)|\phi_{k,d}(t)\rangle = &&\frac{-i\Omega A_{0}}{\Delta}.
\end{eqnarray}
Using the above expressions, we obtain,
\begin{eqnarray}
&&\overline{\dot{\Sigma}^{\rm SS}}=2\beta A_0\Omega {\rm Im}\bigl[\rho_{du}^{\rm SS}\bigr].
\end{eqnarray}
with ${\rm Im}\bigl[\rho_{du}^{\rm SS}\bigr]$ given in Eq.~\eqref{offd3}.

%

\end{document}